\documentclass{aa}  

\usepackage{graphicx}
\usepackage{subcaption}
\usepackage{txfonts}
\usepackage[colorlinks=true, linkcolor=red, urlcolor=cyan, citecolor=blue]{hyperref}
\usepackage{multirow}
\usepackage{color}

\begin{document} 

\title{Gone with the wind: the impact of wind mass transfer on the orbital evolution of AGB binary systems}

\author{M. I. Saladino
          \inst{1,}\inst{2}
          \and
          O. R. Pols\inst{1}
          \and E. van der Helm\inst{2}
          \and I. Pelupessy\inst{3}
	  \and S. Portegies Zwart\inst{2}
          }

\institute{Department of Astrophysics/IMAPP, Radboud University, P.O. Box 9010,
	      6500 GL Nijmegen, The Netherlands\\
              \email{[m.saladino; o.pols]@astro.ru.nl}
         \and
             Leiden Observatory, Leiden University, PO Box 9513, 2300, RA, Leiden, The Netherlands \\
             \email{[vdhelm; spz]@strw.leidenuniv.nl}
             \and
             The Netherlands eScience Center Science Park 140, 1098 XG Amsterdam, The Netherlands\\
             \email{i.pelupessy@esciencecenter.nl}
             }

\date{Received xxxx; accepted xxxx}

\abstract{
In low-mass binary systems, mass transfer is likely to occur via a slow and dense stellar wind when one of the stars is in the asymptotic giant branch (AGB) phase. Observations show that many binaries that have undergone AGB mass transfer have orbital periods of 1-10 yr, at odds with the predictions of binary population synthesis models. In this paper we investigate the mass-accretion efficiency and angular-momentum loss via wind mass transfer in AGB binary systems and we use these quantities to predict the evolution of the orbit. To do so, we perform 3D hydrodynamical simulations of the stellar wind lost by an AGB star in the time-dependent gravitational potential of a binary system, using the AMUSE framework. We approximate the thermal evolution of the gas by imposing a simple effective cooling balance and we vary the orbital separation and the velocity of the stellar wind. We find that for wind velocities larger than the relative orbital velocity of the system the flow is described by the Bondi-Hoyle-Lyttleton approximation and the angular-momentum loss is modest, which leads to an expansion of the orbit. On the other hand, for low wind velocities an accretion disk is formed around the companion and the accretion efficiency as well as the angular-momentum loss are enhanced, implying that the orbit will shrink. We find that the transfer of angular momentum from the binary orbit to the outflowing gas occurs within a few orbital separations from the center of mass of the binary.  Our results suggest that the orbital evolution of AGB binaries can be predicted as a function of the ratio of the terminal wind velocity to the relative orbital velocity of the system, $v_{\infty}/v_\mathrm{orb}$. Our results can provide insight into the puzzling orbital periods of post-AGB binaries and also suggest that the number of stars entering into the common-envelope phase will increase. The latter can have significant implications for the expected formation rates of the end products of low-mass binary evolution, such as cataclysmic binaries, type Ia supernova and double white-dwarf mergers.
}


   \maketitle

\section{Introduction}
\label{sec:introduction}

Depending on their evolutionary state and orbital separation, stars in binary systems can interact in various ways. One of the main interaction processes is the exchange of mass between the stars, which strongly affects the stellar masses, spins and chemical compositions. Furthermore, as a consequence of mass transfer and the loss of mass and angular momentum from the system, significant changes in the orbital parameters can occur \citep[e.g.][]{van_den_heuvel1994,postnov}.
In relatively close binaries mass transfer usually occurs via Roche-lobe overflow (RLOF), while in wide binaries mass transfer by stellar winds can take place. 
Low- and intermediate-mass binary stars usually interact when the most evolved star reaches the red-giant phase or the asymptotic giant branch (AGB). This is a consequence of (1) the large stellar radius expansion during these evolution phases, resulting in a wide range of orbital periods for RLOF to occur, and (2) the strong stellar-wind mass loss of luminous red giants. During the final AGB phase the mass-loss rate is of the order of $10^{-7} - 10^{-4}$ M$_{\odot}$/yr \citep{susanne} and the velocities of the winds are very low (5--30 km/s), which allows for efficient wind mass transfer even in very wide orbits. 

Several classes of relatively unevolved stars show signs of having undergone such mass transfer in the form of enhanced abundances of carbon and $s$-process elements, which are the nucleosynthesis products of AGB stars. These include 
barium (Ba) and CH stars \citep{Keenan1942, Bidelman+Keenan1951}, extrinsic S stars (those in which the radioactive element technetium is absent, \citealp{Smith+Lambert1988}) and carbon-enhanced metal-poor stars enriched in $s$-process elements (CEMP-$s$ stars, \citealp{Beers+Christlieb2005})
These stars are thought to be the result of the evolution of low- and intermediate-mass binaries, that have undergone mass transfer during the AGB phase of the more massive star onto the low mass companion, which is star we currently observe.
The erstwhile AGB star is now a cooling white dwarf (WD) which is in most cases invisible, but which reveals itself by inducing radial-velocity variations of the companion \citep{McClure+1980,McClure1984,Lucatello+2005}.
These systems have orbital periods that lie mostly in the range $10^2 - 10^4$~days and often show modest eccentricities \citep{jorissen+1998,jorissen2,hansen}. They have these orbital properties in common with many other types of binary systems that have interacted during a red-giant phase, such as post-AGB stars in binaries \citep{vanwinckel}, symbiotic binaries \citep{Mikolajewska} and blue stragglers in old open clusters \citep{mathieu}.

These orbital properties are puzzling when we consider the expected consequences of binary interaction during the AGB phase.
Interaction via RLOF from a red giant or AGB star in many cases leads to unstable mass transfer \citep{Hjellming+Webbink1987,Chen+Han2008}, resulting in a common envelope (CE) phase that strongly decreases the size of the orbit \citep{paczynski}. In addition, the orbit is expected to circularise due to tidal effects even before RLOF occurs \citep{Zahn1977,Verbunt+Phinney1995}.
In the case of wind interaction, if the gas escapes isotropically from the donor as is commonly assumed, a fraction of the material will be accreted via the Bondi-Hoyle-Lyttleton \citep[][hereinafter BHL]{Hoyle+Lyttleton1939,Bondi+Hoyle1944} process and the rest of the material escapes, removing some angular momentum and widening the orbit. 
Taking into account these two main mass-transfer mechanisms, binary population synthesis models predict a gap in the orbital period distribution of post-mass-transfer binaries in the range 1--10 yr, and circular orbits below this gap \citep{pols, izzard, nie}. However, the observed period distributions of the various types of post-AGB binaries discussed above show no sign of such a gap; in fact their periods appear to concentrate in the predicted gap. 

This discrepancy points to shortcomings in our understanding of the evolution of binary systems containing red giants and AGB stars. Some of these shortcomings are related to the inadequacy of the BHL accretion scenario to describe the wind interaction process. The BHL description is a good approximation for wind accretion only when the outflow is fast compared to the orbital velocity. These conditions are not met for AGB winds, which are slow compared to the typical orbital velocities and have substantial density gradients. Furthermore, the slow wind velocities allow the escaping gas to interact with the binary and can produce non-isotropic outflows that carry away more angular momentum than in the isotropic case. The complicated interaction between the slow wind and the binary cannot easily be described by an analytical model, and hydrodynamical simulations are needed to study this process.

Several such hydrodynamical studies of the interaction between a red giant undergoing mass transfer via its wind with a companion star have been performed over the last two decades, covering a wide range of phenomena \citep{theuns1,theuns2,mastrodemos,Nagae+2004,shazrene1, val-borro,Val-Borro+2017,huarte,Liu+2017,Chen+2017}. \citet{theuns1} and \citet{theuns2} were the first to carry out 3D smoothed-particle hydrodynamics (SPH) simulations of wind mass transfer, in which they showed that the morphology of the flow differs from BHL expectations. They also showed that the equation of state (EoS) used is important to determine the formation of an accretion disk around the companion. \citet{shazrene1} and \citet{shazrene_thesis} studied the problem of AGB mass transfer in very wide binaries using SPH simulations, in which they took into account the acceleration of the wind by radiation pressure on dust particles. They proposed a new mode of mass transfer, intermediate between RLOF and wind accretion, which they called wind Roche-lobe overflow (WRLOF). They showed that if the dust is formed outside the Roche lobe of the AGB star, wind material can be transferred efficiently through the inner Langrangian point, leading to much higher accretion rates than predicted by the BHL prescription. 

The studies described above have helped to understand the physical processes involved in AGB wind interaction. 
However, only a few studies have focused on the implications for the orbital evolution of binary systems undergoing wind interaction. Given the discrepancy between the timescales that are accessible in simulations and binary evolution timescales, the computation of angular momentum loss is an important way to predict the change in the orbit.
In the context of X-ray binaries, \citet{tavani} developed the first numerical simulations to study the loss of angular momentum and its influence on the binary orbit, by means of ballistic calculations to model wind particles ejected by the mass-losing star. 
Similar calculations are presented in \citet{Hachisu+1999} in the context of wide symbiotic binaries with a mass-losing red giant as potential progenitors of Type Ia supernovae. Both studies find that the specific angular-momentum loss depends on the orbital parameters and the wind ejection velocity.
Although the ballistic treatment and its neglect of any wind acceleration mechanism appears to be adequate for fast winds, for low-velocity mass outflows a hydrodynamical treatment is needed. 
\citet{jahanara2} used a three-dimensional grid-based code to model binary systems with one star undergoing wind mass loss. They study the amount of angular momentum removed as a function of the wind speed at the Roche-lobe surface for different assumed mass-loss mechanisms. Their calculations show that for low wind velocities the specific angular-momentum loss is large (although less so than implied by the ballistic calculations mentioned above), whereas for large velocities the specific angular-momentum loss decreases to the value expected for a non-interacting, isotropic wind. Recently, \citet{rochester} performed grid-based hydrodynamics simulations which also include radiative transfer in order to determine the orbital evolution of binary systems interacting via AGB wind mass transfer. For relatively short orbital periods they find that mass transfer will occur via WRLOF, leading to a shrinking of the orbit and to a possible merger of the components, whereas for wider separations the mass transfer process resembles the BHL scenario and the orbit tends to expand. 

In this paper we try to bridge the previously discussed gap in the orbital period diagram by computing the angular-momentum loss and its effect on the orbit of a binary containing a mass-losing AGB star. To do so, we perform smoothed-particle hydrodynamical simulations including cooling of the gas to model the AGB wind. We present our results as a function of the ratio of the terminal wind velocity to the orbital velocity ($v_{\infty}/v_\mathrm{orb}$). In section \ref{sec:orbit}, we briefly review the equations governing the orbital
evolution of a binary system. In section \ref{sec:method} we describe the model we used and the numerical set-up for the cases we studied, and in section \ref{sec:results} we show the results. In section \ref{sec:discussion} we discuss the implications of our findings for the orbital evolution of binary systems. Finally in section \ref{sec:conclude} we conclude.

\section{Angular momentum loss and mass accretion rate}\label{sec:orbit}

The total orbital angular momentum of a binary system in a circular orbit is given by:
\begin{equation}
\label{total_angular_momentum}
J = \mu a^2 \Omega,
\end{equation}
where $\mu = M_\mathrm{a}M_\mathrm{d}/(M_\mathrm{a}+M_\mathrm{d})$ is the reduced mass of the system, $M_\mathrm{a}$ and $M_\mathrm{d}$ are the masses of the accretor and the donor star respectively, $a$ the orbital separation of the system, and $\Omega$ the angular velocity of the binary. 

If the donor star is losing mass at a rate $\dot{M}_\mathrm{d} <0$ and mass transfer is non-conservative, the companion star will accrete a fraction $\beta$ of the material and the rest will be lost, carrying away angular momentum from the system. The change in orbital angular momentum can be parametrised as:
\begin{equation}
\dot{J} = \eta a^2 \Omega \dot{M}_\mathrm{bin},
\end{equation}
where $\dot{M}_\mathrm{bin} = (1-\beta)\dot{M}_\mathrm{d}$ is the mass-loss rate from the system and $\eta$ is the specific angular momentum lost in units of $J/\mu$. 
Hence, the change in orbital separation for non-conservative mass transfer will be:
\begin{equation}
\label{eq:a}
\frac{\dot{a}}{a} = -2 \frac{\dot{M_\mathrm{d}}}{M_\mathrm{d}} \left[1 - \beta q - \eta(1-\beta)(1+q) - (1-\beta) \frac{q}{2(1+q)}\right],
\end{equation}
where $q=M_\mathrm{d}/M_\mathrm{a}$ is the mass ratio of the stars. This equation can be solved analytically only for a few limiting cases. 
In the Jeans or fast wind mode, mass is assumed to leave the donor star in the form of fast and spherically symmetric wind. Since the speed of the wind
is much larger than the orbital velocity of the system, the wind does not interact with the companion, escaping and taking away the specific orbital angular momentum of the donor ($\eta_\mathrm{iso} = M_\mathrm{a}^2/M_\mathrm{bin}^2$). In the case that $\beta = 0$, the change in the orbit is $\dot{a}/a = -\dot{M}_\mathrm{d}/M_\mathrm{bin}$. This mass transfer mode leads to a widening of the orbit.
The mass transfer efficiency $\beta$ is often described in terms of the BHL analytical model, which can be used as a reference with which to compare the simulation results.
In the framework of a binary system, the BHL accretion rate is given by:
\begin{equation}
\label{rate_bh}
\dot{M}_\mathrm{BHL} = \alpha_\mathrm{BHL}\pi R_\mathrm{BHL}^2 v_\mathrm{rel}\rho,
\end{equation} 
for a high-velocity wind and assuming a supersonic flow. Here, $R_\mathrm{BHL} = 2GM_\mathrm{a}/v_\mathrm{rel}^2$ is the BHL accretion radius, $v_\mathrm{rel} ^2= v_\mathrm{w}^2 + v_\mathrm{orb}^2$ is the relative wind velocity seen by the accretor \citep{theuns2}, $\rho$ is the density at the position of the companion and $\alpha_\mathrm{BHL}$ is the efficiency parameter, of order unity, that physically represents the location of the stagnation point in units of the accretion radius \citep{boffin}.
Using that $\rho = \dot{M}_\mathrm{d}/4\pi a^2 v_\mathrm{w}$ for a steady-state spherical wind and $v_\mathrm{orb}^2 = G(M_\mathrm{a}+M_\mathrm{d})/a$, we can write the accretion efficiency in the BHL approximation as\footnote{In this paper we will refer to this equation as the mass accretion rate predicted by BHL, although for cases in which $v_\mathrm{r} \gg c$, where $c$ is the sound speed, this case corresponds to the Hoyle-Lyttleton approximation.}:
\begin{equation}
\label{BHL}
\beta_\mathrm{BHL} = \frac{\alpha_\mathrm{BHL}}{(1+q)^2}\frac{v_\mathrm{orb}^4}{v_\mathrm{w} (v_\mathrm{w}^2 + v_\mathrm{orb}^2)^{3/2}}.
\end{equation} 

Theoretical considerations and numerical simulations of BHL accretion \citep[e.g. see][]{edgar, matsuda} indicate that for a uniform flow, $\alpha_\mathrm{BHL}$ has a value of about 0.8. In applications to wind accretion Eq. \ref{BHL} is sometimes
divided by a factor of two \citep[e.g.][] {boffin1988,  hurley, carlo1}. 
The corresponding value of $\alpha_\mathrm{BHL}$ is then larger by factor of two \citep[e.g.][use a standard value of $\alpha_\mathrm{BHL} = 1.5$ in their models] {carlo2}.

In realistic situations, and especially in the case of AGB wind mass transfer in binaries, the values of $\beta$ and $\eta$ cannot be expressed in analytical form and have 
to be derived from, for example, hydrodynamical simulations. This is the purpose of this paper. We use the fast wind mode and the BHL accretion efficiency as a reference with
which to compare our simulation results. 

\section{Method}\label{sec:method}

\subsection{Stellar wind}

The mechanism driving the slow winds of AGB stars is not well understood. The pulsation-enhanced dust-driven outflow scenario describes it in two stages \citep{susanne}: in the first stage, pulsations and/or convection cause shock waves which send stellar matter on near-ballistic trajectories, pushing dust-free gas up to a maximum height of a few stellar radii. At this distance, the temperature has dropped enough ($\sim 1500$ K) to allow condensation of gas into dust. In the second stage, the dust-gas mixture is accelerated beyond the escape velocity due to radiation pressure onto the dust. Simulating this process involves several physical mechanisms for which hydrodynamical and radiative transfer codes have to be invoked. This goes beyond the scope of this work. Instead we use a simple approximation to model the velocity profile of the AGB wind. In order to do so, we use the \textsc{stellar\_wind.py} \citep{edwin} routine of the \textsc{amuse}\footnote{\url{http://amusecode.org/}} framework \citep{amuse1,amuse2,amuse3}. Below we briefly explain how this routine works, for a detailed 
description we refer the reader to \citet{edwin}. 

\textsc{stellar\_wind.py} has three different modes to simulate stellar wind. The mode which best approximates wind coming from an AGB star is the accelerating wind mode and it works in the following way. The user provides the stellar parameters, such as mass, effective temperature, radius and mass loss rate, which can be derived from one of the stellar evolution codes currently in \textsc{amuse}. The initial and terminal velocities of the wind, $v_\mathrm{init}$ and $v_{\infty}$, are user input parameters. SPH particles are injected into a shell that extends from the radius of the star to a maximum radius defined by the position that previously released particles have reached after a timestep $\Delta t$. The positions at which these particles are injected correspond to those derived from the density profile of the wind $\rho(r)$. Thus, it is assumed that the velocity and density profiles of the wind, which are related by the mass conservation equation, are known. The wind acceleration function $a_\mathrm{w}(r)$ is obtained from the equation of motion:
\begin{equation}
\label{numerical1}
v\frac{dv}{dr} = a_\mathrm{w}(r) - \frac{GM_{\star}}{r^2} + c^2\left(\frac{2}{r} + \frac{1}{v} \frac{dv}{dr}\right),
\end{equation}
where $v = v(r)$ is the predefined velocity profile for the wind, and $c$ is the adiabatic sound speed in the wind. The second term on the right-hand side describes the deceleration by the gravity of the star and the last term
represents the gas-pressure acceleration. By computing $a_\mathrm{w}$ in this way and applying this acceleration to the gas particles, we guarantee that the gas follows the predefined density profile. If $v_\mathrm{init} = v_\mathrm{\infty}$, then the term on the left hand side will be zero and the acceleration of the wind balances the gravity of the star and the gas pressure gradient. Note that Eq. \ref{numerical1} assumes an adiabatic equation of state (EoS) and ignores cooling of the gas. We have verified that ignoring the cooling term in equation \ref{numerical1} does not change the physical properties of the wind such as density and velocity. The module \textsc{stellar\_wind.py} offers a variety of functions that resemble different types of accelerating winds. Although the most appropriate prescription would be to choose a function with an accelerating region, for the present work we have chosen a constant velocity profile, i.e. $v_\mathrm{init} = v_{\infty}$, for two reasons: to reduce the computational time and to be able to compare our results to other work where similar assumptions have been made.

\subsection{Cooling of the gas}

The internal energy change of the gas plays an important role when modelling the interaction of a star undergoing mass loss with a companion star. \citet{theuns1} found that the formation of an accretion disk around the companion star depends on the EoS. In their study they modelled two cases, one in which the EoS was adiabatic ($\gamma = 5/3$) and one in which the EoS was isothermal ($\gamma = 1$). In the first case, no accretion disk was formed: when material gets close to the companion star, it is compressed by the gravity of the accretor, increasing the temperature and enhancing the pressure, expanding it in the vertical direction. In the isothermal case the pressure does not increase too much, permitting material to stay confined in an accretion disk. Although illustrative, these cases are not realistic and a more physical prescription for cooling of the gas is needed. 

Proper modelling of cooling of the gas requires the implementation of radiative transfer codes. For simplicity, in this study, we use a modified approximation for modelling the change in temperature
in the atmospheres of Mira-like stars based on \citet{bowen}. In his work, the cooling rate $\dot{Q}$ is given by:
\begin{equation}
\label{eq:bowen_cooling}
\dot{Q} = \frac{3k}{2\mu m_{u}} \frac{(T-T_\mathrm{eq})\rho}{C} + \dot{Q}_\mathrm{rad}.
\end{equation}
The first term in Eq. \ref{eq:bowen_cooling} assumes that cooling comes from gas radiating away its thermal energy trying to reach the equilibrium temperature given by the Eddington approximation for a gray spherical atmosphere \citep{chandrasekhar1}:
\begin{equation}
\label{eq:grey_atmosphere}
T^4_\mathrm{eq} = \frac{1}{2} T_\mathrm{eff}^{4} \left\{ \left[1-\left(1-\frac{R_{\star}^2}{r^2}\right)^{1/2}\right] + \frac{3}{2} \int_{r}^{\infty} \frac{R_{\star}^2}{r^2}(\kappa_\mathrm{g} + \kappa_\mathrm{d}) \rho dr\right\}
\end{equation}
where $T_\mathrm{eff}$ is the effective temperature of the star, $\kappa_\mathrm{g}$ and $\kappa_\mathrm{d}$ the opacities of the gas and dust respectively, $R_{\star}$ is the radius of the star, $r$ the distance from the star and $\rho$ the density of the gas. The first term, $W = \frac{1}{2} [1-[1-R_{\star}^2/r^2)^{1/2}],$ corresponds to the geometrical dilution factor, and the term in the integral plays the role of the optical depth in plane-parallel geometry. The constant $C$ in equation \ref{eq:bowen_cooling} is a parameter reflecting the radiative equilibrium timescale. Following \citet{bowen}, we adopt a value of  $C = 10^{-5}$ g s cm$^{-3}$.

The second term in equation \ref{eq:bowen_cooling} corresponds to radiation losses for temperatures above $\sim 7000$ K, where the excitation of the $n=2$ level of neutral hydrogen is mainly responsible for the energy loss \citep{spitzer}. In this work however, we use an updated cooling rate prescription for high temperatures ($\log T \geq 3.8$ K) by \citet{schure}. The advantage of this prescription is that contributions to the cooling rate come not only from neutral hydrogen but individual elements are taken into account according to the abundance requirements. We use the abundances for solar metallicity given by \citet{andersandgrevesse}. Then, the second term in equation \ref{eq:bowen_cooling} becomes:
\begin{equation}
\label{eq:cooling_rate_main}
\dot{Q}_\mathrm{rad} = \frac{\Lambda_\mathrm{hd} n_\mathrm{H}^2}{\rho}
\end{equation}
with $\Lambda_\mathrm{hd}$ interpolated between the values given in table 2 of \citet{schure} and $n_\mathrm{H} = X\rho/m_\mathrm{H}$. 

For low temperatures the first term of equation \ref{eq:bowen_cooling} is the dominant term. If this term is not taken into account the internal energy of the particles reaches very low values leading to unphysical temperatures. Note that in equation \ref{eq:grey_atmosphere}, the optical depth at large distances from the star will be very small regardless of the opacity values, thus for distant regions the equilibrium temperature will be mainly determined by the geometrical dilution factor $W$. For this reason, and since the opacities are only used to calculate the equilibrium temperature at given radius, the opacities in equation \ref{eq:grey_atmosphere} are taken constant during the simulations with values of $\kappa_\mathrm{g} = 2 \times 10^{-4}$ cm$^{2}$ g$^{-1}$ and $\kappa_\mathrm{d} = 5$ cm$^{2}$ g$^{-1}$ \citep{bowen}. 


\subsection{Computational method}

To calculate self-consistently the 3D gas dynamics of the wind in the evolving potential of the binary, we used the SPH code \textsc{fi} \citep{fi1,fi2,fi3} with artificial viscosity parameters as shown in Table \ref{table:numerical_parameters}.  The AGB star was modelled as a point mass, whereas the companion star was modelled as a sink particle with radius corresponding to a fraction of its Roche Lobe radius $R_\mathrm{L,a}$. In order to model the dynamics of the stars, we coupled the SPH code with the N-body code \textsc{huayno} \citep{huayno} using the \textsc{bridge} module \citep{bridge} in \textsc{amuse}. The N-body code is used to evolve the stellar orbits, and to determine the time-dependent gravitational potential in which the gas dynamics is calculated. Since the interest in this work is to study the evolution of the gas under the influence of the binary system, the \textsc{bridge} module allows the gas to feel the gravitational field of the stars, however the stars do not feel the gravitational field of the gas; also self-gravity of the gas is neglected. This is a fair assumption given that the total mass of the gas particles in the simulation ($4\times 10^{-5}$ M$_{\odot}$)  is very small compared to the binary mass (4.5 M$_{\odot}$). This assumption does not have a significant effect on the evolution of the gas, as was verified in a test simulation. 

\begin{table}[h]
\caption{Fixed parameters in the simulations}
\label{table:numerical_parameters}
\begin{tabular}{l l l}
\hline\hline
Parameter	&	Value	&	Description	\\ \hline
$M_\mathrm{d}$	&	3 M$_{\odot}$	&	Mass of donor star	\\
$M_\mathrm{a}$	&	1.5 M$_{\odot}$	&	Mass of accretor	\\
$R_\mathrm{d}$	&	200 R$_{\odot}$	&	Inner boundary for SPH\\
 & & particles about donor star \\

$\dot{M}_\mathrm{d}$	&	$10^{-6}$ M$_{\odot}$ yr$^{-1}$	&	Mass loss rate 	\\
$\alpha_\mathrm{SPH}$	&	0.5	&	Artificial viscosity parameter	\\
$\beta_\mathrm{SPH}$	&	1	&	Artificial viscosity parameter	\\ \hline\hline
\end{tabular}
\end{table}

\subsubsection{Binary set up}

\begin{table*}[h]
\centering
\caption{Simulation parameters. The values of $M_\mathrm{a}$, $M_\mathrm{d}$ and $\dot{M}_\mathrm{d}$ are the same for all simulations.}
\label{table1}
\begin{tabular}{c c c c c c c c c c} 
\hline\hline
	&	\# &	$a$	&	$v_{\infty}$	&	$v_{\infty}/v_\mathrm{orb}$	& $R_\mathrm{a}$	& Cooling	&	$m_\mathrm{g}$	&	$\bar{h}$ &$n$	\\ \hline \hline
\multirow{3}{*}{Test models}	&	T1	&	3	&	15	&	0.41	&  20.8 	& Adiabatic	&	$10^{-10}$&  0.24 (0.16--0.40)	&	$2\times 10^5$	\\
	&	T2	&	3	&	15	&	0.41	& 20.8	& Isothermal	&	$10^{-10}$ & 0.12 (0.01--0.36)	&	$2\times10^5$	\\
	&	T3	&	3	&	15	&	0.41	& 20.8	& Bowen + Schure	&	$10^{-10}$ & 0.13 (0.01--0.37)	&	2$\times10^5$	\\ \hline
\multirow{3}{*}{Test resolution}	&	R1	&	5	&	15	&	0.53	&  34.7	& Bowen + Schure &  $10^{-11}$ 	& 0.13 (0.01--0.25)	&$3.3\times 10^6$	\\
	&	R2	&	5	&	15	&	0.53	& 34.7	& Bowen + Schure	&	$4 \times 10^{-11}$	& 0.20 (0.02--0.39)    & $1.1\times 10^6$	\\
	&	R3	&	5	&	15	&	0.53	& 34.7	& Bowen + Schure	&	$8 \times 10^{-11}$	&	0.26 (0.03--0.49) & $5.6\times 10^5$	\\ \hline
\multirow{4}{*}{Science models}	&	V10a5	&	5	&	10	&	0.35	&  34.7	& Bowen + Schure	&	$2 \times 10^{-11}$	& 0.17 (0.06--0.29)	& $2.2\times10^6$	\\
 	&	V15a5	&	5	&	15	&	0.53	&  34.7	& Bowen + Schure	&	$2 \times 10^{-11}$	& 0.16 (0.01--0.32) & $2.2\times10^6$	\\
	&      V30a5	&	5	&	30	&	1.06	&   34.7	& Bowen + Schure	&	$2 \times 10^{-11}$	&	0.29 (0.14--0.45) & $2.2\times10^6$	\\
	&	V150a5 	&	5	&	150	&	5.31	&  34.7	& Bowen + Schure	&	$2 \times 10^{-11}$	&	0.50 (0.27--0.78) & $2.2\times10^6$	\\
	&	V19a3	&	3	&	19.33	&	0.53	& 20.8	& Bowen + Schure	&	$2 \times 10^{-11}$	&	0.07 (0.004--0.22) & $10^6$	\\ \hline
\multirow{2}{*}{Test sink radius} 
        &V10a5s2 &  5       &     10     &      0.35 &   20.8   & Bowen + Schure       &      $8 \times 10^{-11}$   & 0.14 (0.01--0.43)   &  $5.6 \times10^5$ \\
        & V15a5s2 & 5 & 15 & 0.53 & 20.8 & Bowen + Schure      &  $2 \times 10^{-11}$    & 0.08 (0.004--0.28)  & $2.2\times10^6$	\\
	&     V30a5s2 &     5      &      30     &      1.06   &  20.8    & Bowen + Schure       &     $2 \times 10^{-11}$	&	0.28 (0.14--0.45) & $2.2\times10^6$	\\	\hline \hline
        
\end{tabular}
\tablefoot{The second column corresponds to the name of the model.
$a$ is the initial orbital separation of the system in AU.
$v_{\infty}$ is the assumed terminal velocity of the wind in km s$^{-1}$.
The ratio $v_{\infty}/v_\mathrm{orb}$ is the ratio of the terminal velocity to the orbital velocity of the system.
$R_{a}$ is the radius of the sink in R$_{\odot}$.
Column 7 specifies the EoS for the model. 
$m_\mathrm{g}$ corresponds to the mass of the SPH particles in units of M$_{\odot}$.
$\bar{h}$ corresponds to the mean smoothing length of the SPH particles within radius $a$ from the center of mass of the system in units of AU. 
The numbers brackets correspond to the 95\% confidence interval, i.e. 2.5\% of the particles have values below the lower limit and 2.5\% 
particles have values above the upper limit of the given interval.
The final column gives the total number of particles $n$ generated at the end of the simulation.}
\end{table*} 

The stellar parameters of our simulated systems are chosen to match those presented in \citet{theuns1}. This will allow a direct comparison between our results and their work. The mass of the AGB star is $M_\mathrm{d}$ = 3 M$_{\odot}$ and that of the low-mass companion is $M_\mathrm{a}$ = 1.5 M$_{\odot}$. The radius of the primary star is $R_\mathrm{d}$ = 200 R$_{\odot}$ and for the secondary we assume a sink radius equal to $R_\mathrm{a}$ = $0.1 R_\mathrm{L,a}$. The orbit is circular and the separation of the stars is $a = 3$ AU in our test simulations. The mass-loss rate $\dot{M_\mathrm{d}} = 10^{-6}$ M$_{\odot}$ yr$^{-1}$ is constant during the simulation; the terminal velocity of the wind is $v_{\infty} = 15$ km s$^{-1}$ and the gas is assumed to be monoatomic $\gamma = 5/3$ with constant mean molecular weight $\mu = 1.29$ corresponding to an atomic gas with solar chemical composition ($X = 0.707, Y= 0.274, Z  = 0.019$) \citep{andersandgrevesse}. The effective temperature of the AGB star, which is also the initial temperature given to the gas particles, is $T_\mathrm{eff} = 2430$ K for all models except the isothermal test case, for which $T_\mathrm{eff} = 4050$ K in order to compare to \citet{theuns2}. We assume the AGB star to be non-rotating as a result of its history of expansion after the main sequence, combined with angular momentum loss resulting from earlier mass loss on the RGB and AGB. Thus, we ignore the possibility of subsequent spin-up by tidal interaction (see Sect. \ref{sec:rotation} for a discussion). The stars are placed in such a way that the centre of mass is located at the origin of  the system of reference. The setup for the other simulations is the same except for the values of the separations and the terminal velocities of the wind. We also ran a few simulations with a smaller sink radius (see Table \ref{table1}). Each simulation was run for eight orbital periods. 

\subsection{Testing the method}\label{testing}

\begin{figure*}
\centering
\begin{subfigure}{1.0\textwidth}
   \includegraphics[width=1\linewidth]{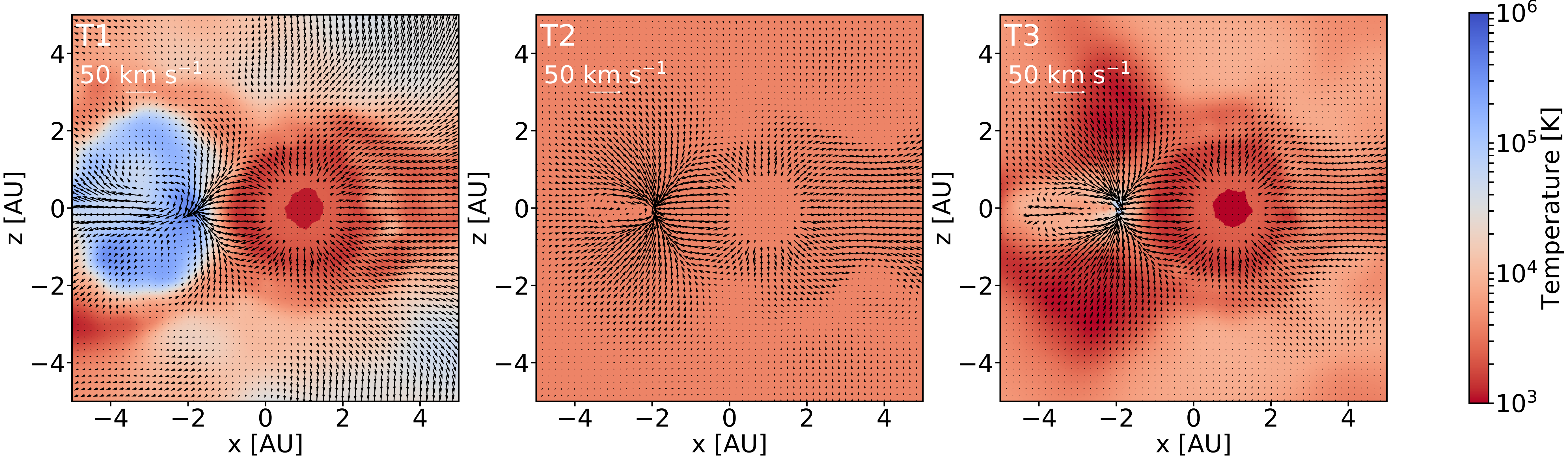}
\end{subfigure}

\begin{subfigure}{1.0\textwidth}
   \includegraphics[width=1.02\linewidth]{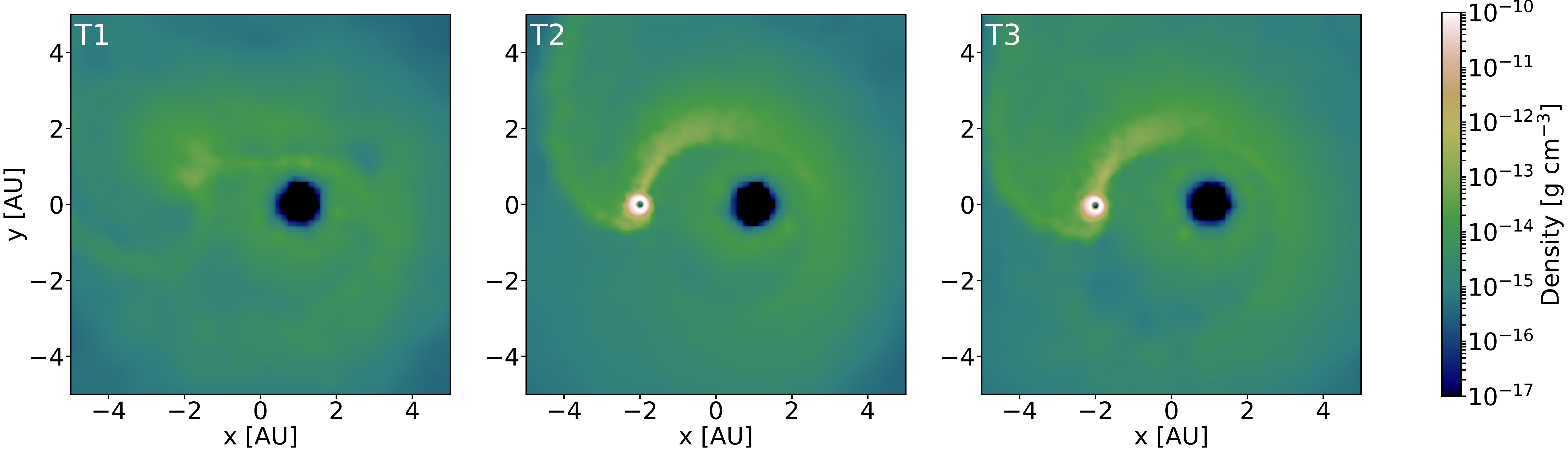}
\end{subfigure}
\caption{\emph{Top:} Flow structure in the $y=0$ plane for different EoS after 7.5 orbits. The colormap shows the temperature at $y=0$. The white arrow on top of the velocity field plots corresponds to the magnitude of a 50 km s$^{-1}$ velocity vector. In the three images we can observe the
wind leaving the donor star radially at $x = 1$ AU. The accretor star is located at $x=-2$ AU. For model T1, the temperatures reached in the region near the companion star are very high preventing material to settle into an accretion disk. 
\emph{Bottom:} The gas density in the orbital plane ($z=0$) is shown. The accretor is located at $x = -2, y =0$, and the AGB star at $x=1, y=0$. In the middle panel we see that for the isothermal EoS (T2) gas in the vicinity of the accretor settles into an accretion disk. When cooling is included (T3), an accretion disk around the companion is also formed.}
\label{fig:iso_ad_cool}
\end{figure*}

In this subsection, we compare our simulations (models T1, T2, T3 in Table \ref{table1}) with those of \citet{theuns1}. The top panel of Figure \ref{fig:iso_ad_cool} shows the velocity field of the gas in the $y=0$ plane, perpendicular to the orbital plane, for models T1, T2 and T3 after 7.5 orbits. The colormap shows the temperature of the gas in the $y=0$ plane. The donor star is located at $x = 1$ AU, $z=0$ AU and the companion star at $x=-2$ AU, $z=0$ AU. Gas expands radially away from the primary star. Similar to \citet{theuns1}, in the adiabatic case (model T1) we observe 
a bow shock close to the companion star where high temperatures are reached. The high-temperature regions are located symmetrically above and below the star in this plane. The gravitational compression of the gas by the companion also enhances the temperature
in this region, expanding the gas in the vertical direction and preventing it from creating an accretion disk. On the other hand, when an isothermal equation of state is used (model T2), an accretion disk is formed around the companion. In addition, two spiral arms are observed around both systems similar to those observed by \citet{theuns1}. These features can be seen in the bottom panel of Figure \ref{fig:iso_ad_cool}, where density in the orbital plane ($z=0$) is shown.
Model T3 shows that using an EoS that includes cooling of the gas (Equation \ref{eq:bowen_cooling}) gives a similar outflow structure as in the isothermal case, including the 
formation of an accretion disk. However, in this model there is an increase in temperature where the spiral shocks are formed. These effects are also found in our science simulations, which we describe in detail in section \ref{science}.

\subsection{Mass accretion rate and angular momentum loss}\label{sec:steady}

We measure the mass accretion rate by adding the masses $m_\mathrm{g}$ of the gas particles that entered the sink during each timestep $\delta t$ in the simulation. Due to the discreteness of the SPH particles, the resulting accretion rates will be subject to shot noise. This can be seen as fluctuations in the mass accreted as a function of time. In order to suppress statistical errors in the results, we average the mass accretion rate over longer intervals of time.

Measuring the angular momentum loss rate from the system is more complicated. An advantage of SPH codes is that they conserve angular momentum extremely well compared to Eulerian schemes \citep{price_mother}. One question that arises is at which distance an SPH particle is no longer influenced by the gravitational potential of the binary system. In order to determine this, we construct spherical shells at various radii $r_\mathrm{S}$ from the center of mass of the binary. In the same way as for the accretion rate, we compute the net mass-loss rate through such a surface. Each particle $i$ carries angular momentum $\vec{J}_i$, of which we add up the components $J_{z,i}$ along the $z$-axis, perpendicular to the orbital plane. We compute the specific angular momentum of the outflowing mass as follows,
\begin{equation}
\label{eq:j}
j_\mathrm{loss} = \frac{\sum_i^N J_{z,i}}{\Delta N m_\mathrm{g}},
\end{equation}
with $\Delta N$ the number of particles that crossed the shell during a timestep $\Delta t$. Similar to the accretion rate, we average the mass and angular-momentum loss over longer intervals of time to suppress statistical fluctuations. We also verified that, once the simulation has reached a quasi-steady state, the orthogonal ($x$ and $y$) components of the angular momentum-loss are negligible.

\begin{figure}
\centering
\includegraphics[width=\hsize]{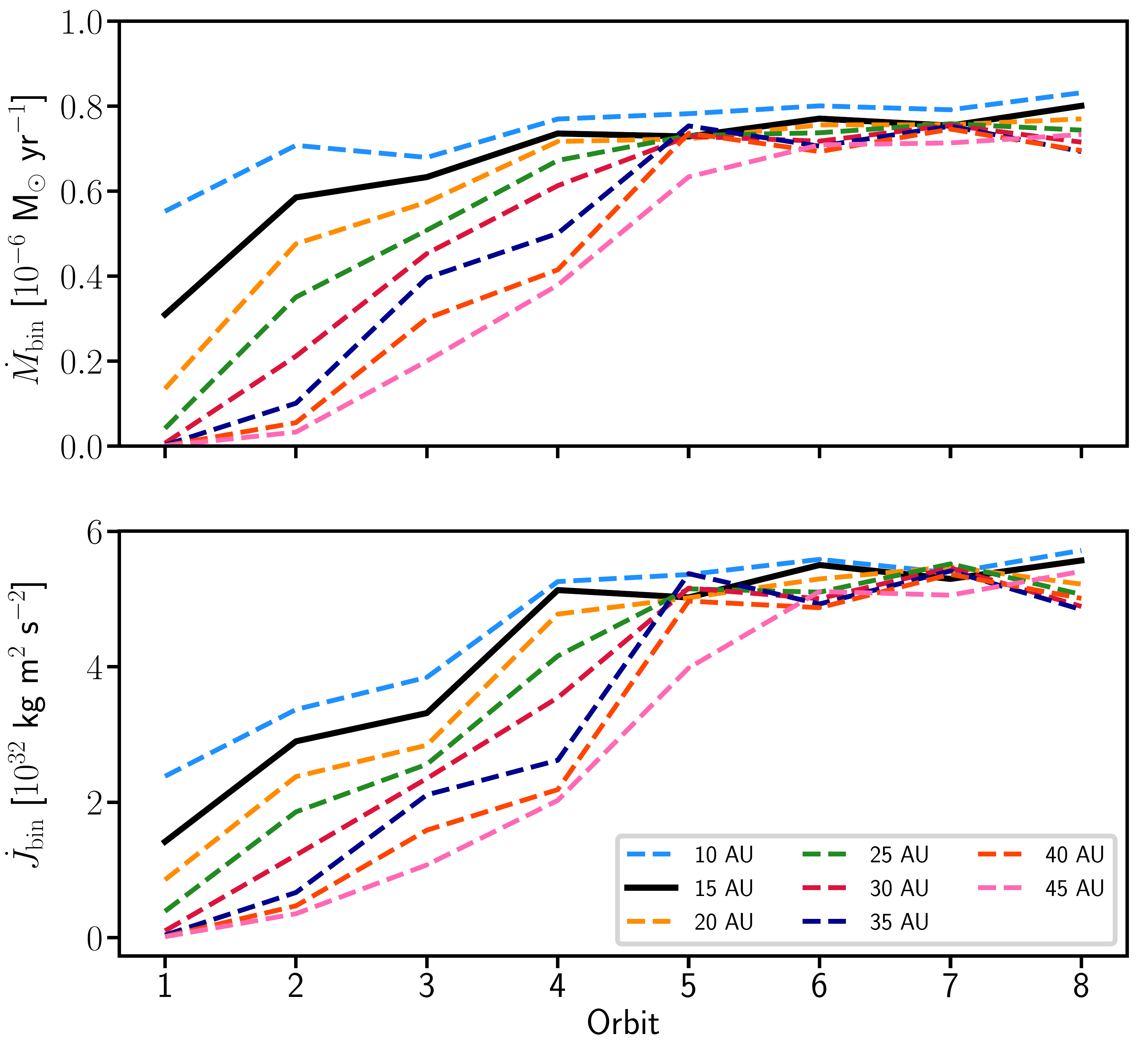}
\caption{In the top panel we show the average mass loss per orbit for the standard model V15a5 measured
at different radii from the center of mass of the binary as indicated by different colours.  
In the bottom panel the corresponding angular-momentum loss rate is shown. 
The black thick curve for $r_\mathrm{S} = 15$ AU corresponds to the radius at which we measure $\dot{J}_\mathrm{bin}$.}
\label{fig:mass_loss}
\end{figure}

In Figure \ref{fig:mass_loss} we show the resulting average mass and angular momentum-loss per orbit in our standard model V15a5 (see Table \ref{table1}) as a function of time for different shell radii $r_\mathrm{S}$. After several orbits, both quantities become approximately constant in time and almost independent of the chosen shell radius. This steady state is reached later for larger radii, corresponding to the longer travel time of the wind particles to reach this distance from the donor star. The near-constancy of the angular-momentum loss rate as a function of $r_\mathrm{S}$ indicates that the torque transferring angular momentum from the orbit to the escaping gas operates inside a few times the orbital separation $a$. Beyond $r_\mathrm{S} \approx 10$ AU ($\approx 2a$) the transfer of angular momentum to the gas is essentially complete, and angular momentum is simply advected outwards with the flow. We therefore adopt the the mass and angular momentum loss values measured at $r_\mathrm{S} = 3a$, which reach a steady state after about 4 orbits in model V15a5. The time taken to reach the quasi- steady state depends on the terminal velocity of the wind, being shorter for larger velocity. 

Once a simulation has reached a quasi-steady state, we compute the mass accretion efficiency $\beta = \dot{M}_\mathrm{a}/\dot{M}_\mathrm{d}$ from the average mass accretion rate over the remaining $N$ orbits in the simulation. Similarly, we average the angular momentum loss (Eq. \ref{eq:j}) over the last $N$ orbits to compute the value of $\eta$, i.e.\ the specific angular momentum of the lost mass in units of $a^2 \Omega$.

\section{Results}\label{sec:results}

We performed 13 different simulations, listed in Table \ref{table1}, for which we discuss the results in the following subsections. The first three simulations (T1-T3) are performed to test the EoS (see Sect. \ref{testing}). The following 
three simulations (R1-R3) correspond to the test of different SPH resolutions, compared to simulation V15a5 which we consider the standard model. For this model we chose an orbital separation of 5 AU and a terminal velocity of the wind of 15 km s$^{-1}$. V10a5, V30a5 and V150a5 correspond to simulations with the same orbital separation but different velocities of the wind, either larger or smaller than for the standard case. V10a5s2, V15a5s2 and V30a5s2 correspond to
test simulations in which we study the effect of assuming a smaller sink radius ($R_\mathrm{a} = 0.06 R_\mathrm{L, a}$) on the mass-accretion rate and radius of the accretion disk. Note that simulation V10a5s2 was performed at a lower resolution than the others. Finally in V19a3 we change the orbital separation to 3 AU, keeping the same ratio of the velocity of the wind to the orbital velocity of the system as in V15a5. 

\subsection{Convergence test}\label{sec:convergence}

Choosing the best resolution for SPH simulations is not simple and depends on the physical process under investigation. To determine the optimal resolution within a reasonable computational time, we performed four simulations of our standard model in which we vary the mass of the SPH particles $m_\mathrm{g}$ (R1, R2, R3 and V15a5 in Table \ref{table1}). Note that a different $m_\mathrm{g}$ also implies a change in the
average smoothing length $\bar{h}$ of the SPH particles. Table \ref{table1} shows the value of $\bar{h}$ within a radius $a$ of the center of mass of the binary, and the 95\% confidence interval of smoothing-length values within this radius. The total number of particles $n$ generated during one simulation 
is given by $n = \dot{M}_\mathrm{d} t_\mathrm{end}/m_\mathrm{g}$. Thus, the number of particles generated for the lowest resolution (largest $m_\mathrm{g}$) is $n = 5.6 \times 10^{5}$ and $n = 3.3 \times 10^{6}$ for the highest resolution (smallest $m_\mathrm{g}$). This last simulation ran only for 6 orbital periods.  

\begin{figure}[h]
\centering
\includegraphics[width=\hsize]{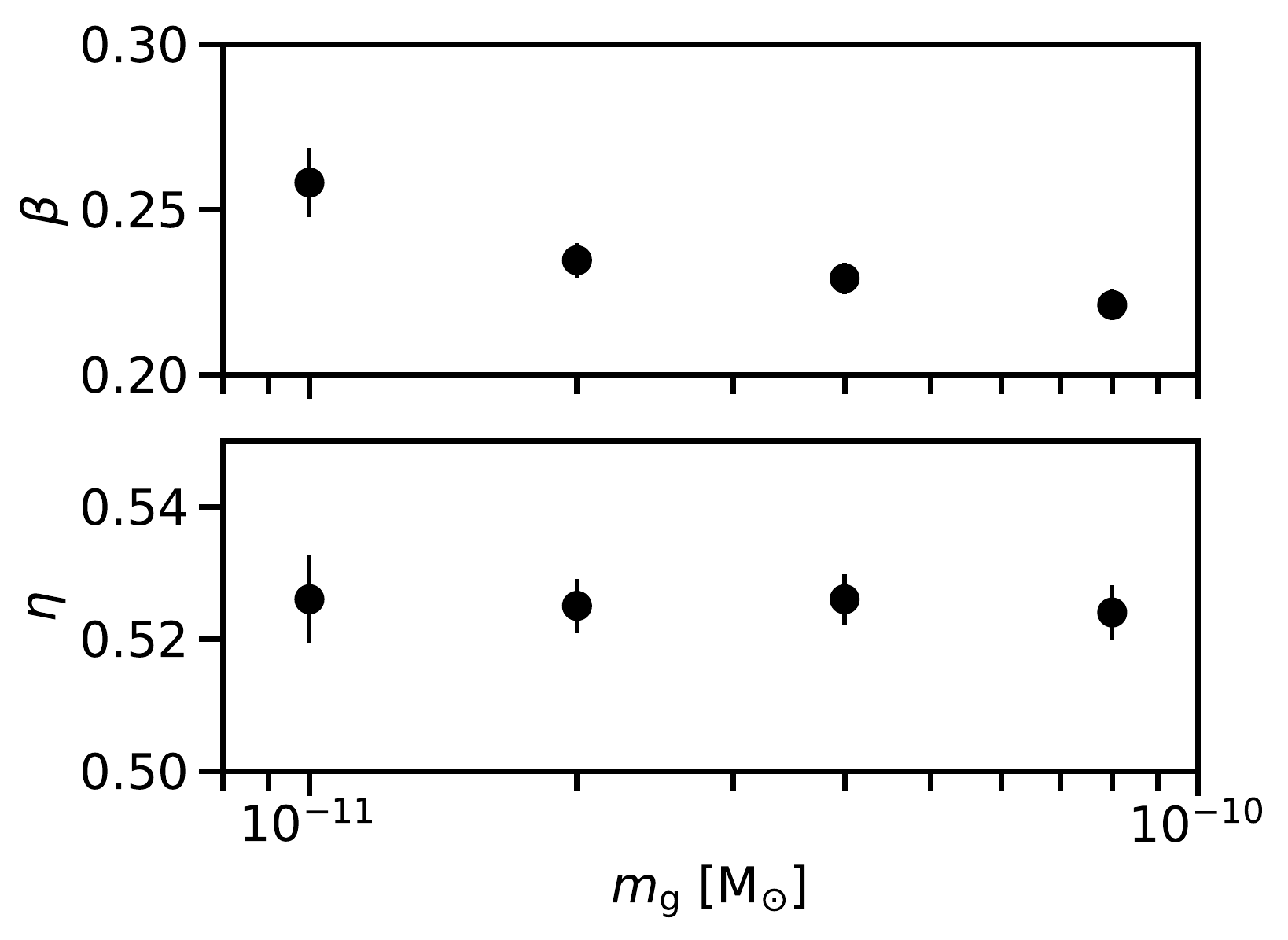}
\caption{The mean values of the $\beta$ and $\eta$, relating to mass and angular momentum loss, as a function of the mass of the SPH particles used to test different resolutions for the simulations. The error bars correspond to the overall standard error of the mean $\sigma_\mathrm{m}$ multiplied by a factor of 5 for better appreciation.}
\label{fig:convergence}
\end{figure}

Since the interest of this study is to obtain numerical estimates for the average mass-accretion efficiency and average angular-momentum loss during the simulation, these were the quantities we used to check for convergence. Figure \ref{fig:convergence} shows the values obtained for these quantities (as explained in Section \ref{sec:steady}) as a function of the mass of the SPH particles $m_\mathrm{g}$. The error bars in this figure correspond to five times the standard error of the mean $\sigma_\mathrm{m}$, which was estimated by means of $\sigma_\mathrm{m}^2 = 1/N \sum_{i=1}^{N} \sigma_{\mathrm{m, orb}_i}^2$, where $\sigma_{\mathrm{m, orb}_{i}}$ is the standard error of the mean per orbit\footnote{The
standard error of the mean per orbit is: $$\sigma_\mathrm{m, orb} = \frac{1}{N_p} \sqrt{\sum_{i=1}^{N_p} (\dot{m}_i - \left<\dot{m}\right>)^2},$$ where $N_{p}$ is the number of timesteps ($t_1, t_2, ..., t_p$) into which the orbit is divided, $\dot{m}_i$ the mass accretion rate at time $t_i$ and $\left<\dot{m}\right>$ the average mass accretion rate per orbit.}
and $N$ the number of orbits during which the quasi-steady state has been reached. Since the number of timesteps over which the average is taken is large, the standard error turns out to be very small. This is the reason why we chose to plot five times the value of $\sigma_\mathrm{m}$ in Figure \ref{fig:convergence}.  
Because the simulation with the highest resolution (smallest $m_\mathrm{g}$) only ran for 6 orbital periods, and the system reaches the quasi-steady state after 4 orbits, the error in the mean is larger than for the other simulations. 

One thing that can be seen from Figure \ref{fig:convergence} is that, even though by only a small amount, the accretion efficiency increases with increasing resolution, contrary to what \citet{theuns2} observed. However, we note that in their work the accretion rate was computed by setting an accretion radius proportional to the resolution of the SPH particles (in terms of the smoothing length $h$), whereas in our case the radius of the sink has the same constant value for all our models. On the other hand, the angular momentum per unit mass lost by the system is approximately independent of the resolution. 

We also find that the time variability in the mass-accretion rate (described in Sect. \ref{sec:mass_accretion_rate}) increases
with increasing resolution. It is not clear why this takes place, but we should bear in mind that this variability could also be an artefact of the numerical method.
As will be discussed in Sect. \ref{sec:mass_accretion_rate} there appears to be a correlation between the accretion disk mass and the mass-accretion rate. Given that the results of SPH simulations
for accretion disks depend on the formulation of artificial viscosity, we performed a test simulation with the lowest  resolution ($m_\mathrm{g} = 8\times10^{-11}$) adopting artificial viscosity parameters in \textsc{Fi} equal to those used by \citet{wijnen1} for their protoplanetary disk models ($\alpha_\mathrm{SPH} = 0.1$ and $\beta_\mathrm{SPH} = 1$).
We find almost the same value for the average mass-accretion rate ($\beta = 0.223 \pm 0.002$, versus $\beta = 0.221 \pm 0.002$ for simulation R3). With this alternative prescription of artificial viscosity we find similar variability in the mass-accretion rate.

In order to guarantee reliable results within a reasonable computational time, the chosen resolution for the other simulations is the one in which the mass of the individual SPH particles is $m_\mathrm{g} = 2\times10^{-11}$ M$_{\odot}$. For this simulation, the value of $\beta$ differs by 9\% compared to the value obtained for the simulation with largest resolution (R1). This is indicative of the numerical accuracy of the accretion rates obtained in our simulation. By contrast, the angular-momentum loss rates ($\eta$ values) we measure differ by <0.4\% between simulations at different resolution.

\subsection{Summary of model results}\label{science}

\begin{table*}[h]
\centering
\caption{Simulation results and orbital evolution parameters}
\label{table:main}
\begin{tabular}{c c c c c c c c c c} 
\hline\hline
Simulation	&	$N$	&	$\beta_\mathrm{sink}$	& $\beta_{0.4R_\mathrm{L,2}}$	&	$\beta_\mathrm{BHL}$ &		$\eta$			&	Disk	&	$R_\mathrm{disk}$	& $M_\mathrm{disk}$	&$\dot{a}/a$ \\ \hline
T1    &      4       &    0.0291    &  0.0183          &     0.21         &  0.452  &  No &    -  & - & $-4.65 \times 10^{-7}$\\
T2    &      4       &  0.289   &  0.348  &   0.21    &  0.623  &  Yes & $\approx 0.37$  & $ \approx 1.3 \times 10^{-6}$ & $-7.54 \times 10^{-7}$  \\
T3	&	4	        &	0.309	& 	0.341	&  0.21  &	$	0.681		$	&	Yes	&	$\approx 0.37$ 	 & $\approx 1.0\times10^{-6}$	&$-8.32 \times 10^{-7}$		\\
V10a5	&	4	&	0.3879 & 0.366 	& 0.26 &	$	0.641		$	&	Yes	&	$\approx 0.40$	& $\approx 2.4\times10^{-7}$	&$-7.70 \times 10^{-7}$		\\
V10a5s2   &      4  &  0.344    &  0.388    & 0.26 &      $       0.609    $      &      Yes   &     $\approx 0.40$          &  $\approx 2.8\times10^{-6}$ &  $-7.33 \times 10^{-7}$        \\
V15a5	&	4	&	0.235 &  0.231 	&0.14 &	$	0.525		$	&	Yes	&	$\approx 0.47$ & $\approx 3.6\times10^{-7}$	&	$-6.20 \times 10^{-7}$		\\
V15as2     &      4   &  0.1563    & 0.237   & 0.14    &    $       0.5315           $      &       Yes  &     $\approx 0.56$ & $\approx 3.3\times10^{-6}$ &   $-6.30 \times 10^{-7}$ \\
V19a3	&	4	&   0.260  &	0.316	&0.14&	$	0.544		$	&	Yes	&	$\approx 0.30$	 & $\approx 1.3 \times10^{-6}$	&$-6.51 \times 10^{-7}$		\\ 
V30a5	&	6	& 0.039  & 0.039	 &  0.037 &	$0.1849$	&	No	&	-	&   -	&$6.06 \times 10^{-8}$	 		\\
V30a5s2   &  6  & 0.0353  &    0.0353   & 0.037&    $0.1862 $       &     No    &      -       & -  &    $4.35 \times 10^{-8}$           \\
V150	a5 &	7	&	$	3.46 \times 10^{-4} $ &  $3.45\times 10^{-4}$	& $1.32 \times 10^{-4}$ &	$	0.1116	$	&	No	&	-	&  -	&$2.21 \times 10^{-7}$		\\
\hline\hline \\
\end{tabular}
\tablefoot{$N$ is the number of orbits over which the values are averaged after the quasi-steady state is reached. 
$\beta_\mathrm{sink}$ is the averaged mas-accretion efficiency measured from the net sink inflow.
$\beta_{0.4R_\mathrm{L,2}}$ is the averaged mass-accretion efficiency measured from the net inflow into a shell of radius 0.4$R_\mathrm{L,2}$, i.e. it includes the mass inflow of the sink and of the accretion disk, if present. 
$\beta_\mathrm{BHL}$ corresponds to the mass accretion efficiency as predicted by BHL (Eq. \ref{BHL}) with $\alpha_\mathrm{BHL} = 1$.
$\eta$ is the averaged specific angular momentum of material lost.
$R_\mathrm{disk}$ is the approximate radius of the disk in AU after eight orbital periods, except for model V10a5, where the radius corresponds to 2 orbital periods.
$M_\mathrm{disk}$ is the mass within $R_\mathrm{disk}$.
$\dot{a}/a$ is the relative rate of change of the orbit as derived from Equation \ref{eq:a} in units of yr$^{-1}$. Positive sign means the orbit is expanding, whereas negative sign means that the orbit is shrinking.
}
\end{table*}

\subsubsection{Morphology of the outflow}\label{morphology}

\begin{figure*}
\centering
\includegraphics[width=0.9\textwidth]{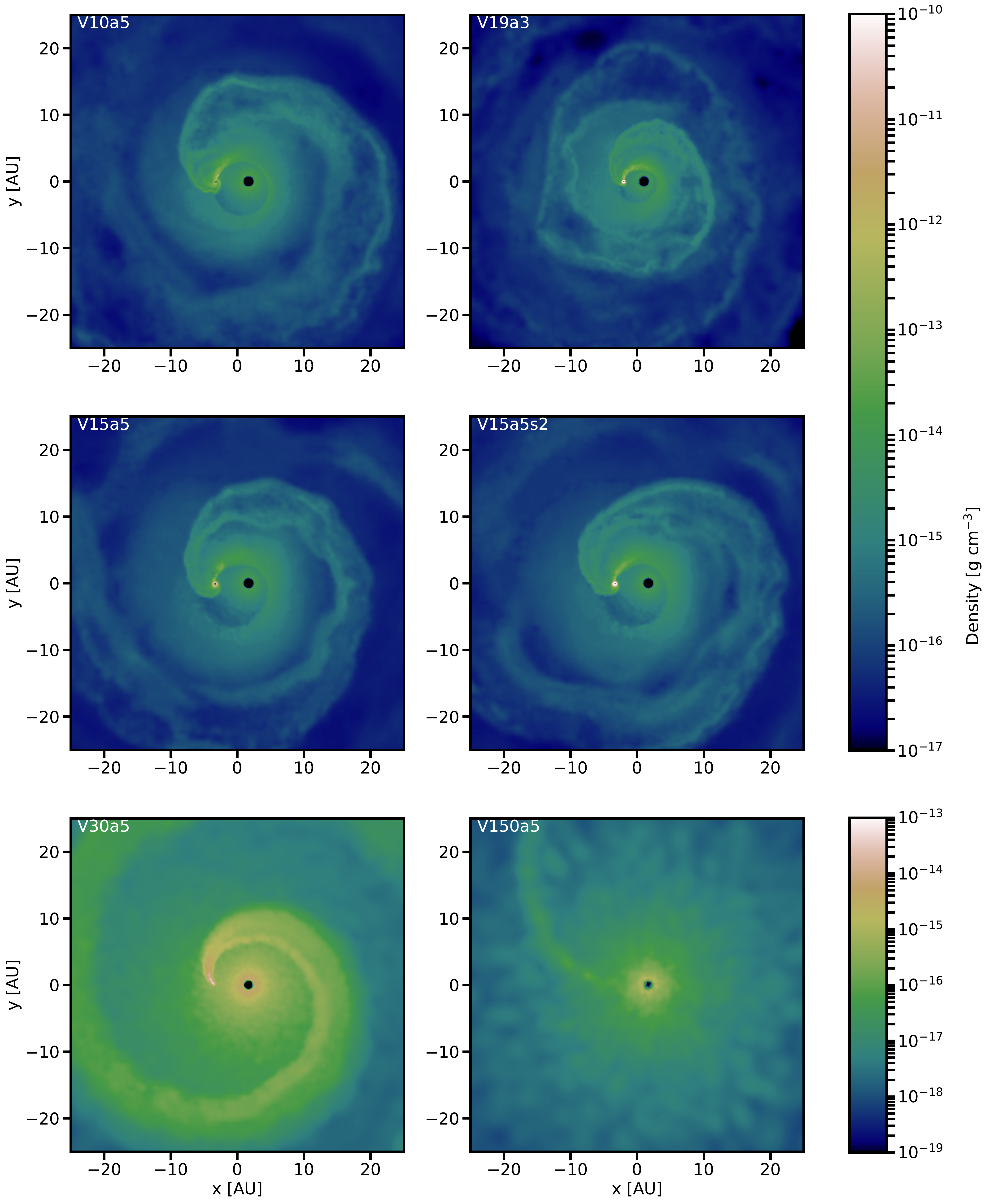}
\caption{Gas density in the orbital plane ($z=0$) for V10a5, V10a5s2, V15a5, V19a3, V30a5 and V150a5 after 7.5 orbits.}
\label{fig:overall}
\end{figure*}

Since the velocity of the wind is of the same order as the orbital velocity of the system, Coriolis effects play an important role in shaping the outflow. 
Figure \ref{fig:overall} shows the density of the gas in the orbital plane for different simulations after 7.5 orbits. 
The orbital motion is counterclockwise in this view.
We observe that in the simulations with relatively low wind velocities, $v_\mathrm{\infty}/v_\mathrm{orb} < 1$ (V10a5, V15a5, and V19a3), two spiral arms are formed around the system. These spiral arms delimit the accretion wake of the companion star. 
Similar to \citet{theuns1}, we find that the inner spiral shock, wrapped around the donor star, is formed by stellar-wind material leaving the AGB star colliding with gas moving away 
from the companion star with $v_{x} >0$.
Near the companion star a bow shock is created by the wind coming from the AGB star as it approaches the accretor.
The temperature of the gas in this region increases and material is deflected behind the star forming the outer spiral arm. 
The Mach number of the flow near the accretor has values between 2 and 6.
Because of its proximity to the donor star, where the gas density is high, the inner spiral arm has higher densities than the outer spiral arm. In these systems we also observe an accretion disk around the companion (Sect. \ref{sec:accretion_disk}). The disk is asymmetric and is rotating counterclockwise.  

Figure \ref{fig:overall} shows that for model V30a5 the density of the gas in the accretion wake is lower than in models with $v_{\infty}<v_\mathrm{orb}$ (note that a different colour scale is used in this panel). The density of the accetion wake is high close to the companion star, but it decreases along the stream. 
Two spiral arms delineate the edge of the accretion wake, which converge into what appears to be a single arm close to the companion star.
No accretion disk is observed in this model. For model V150a5 a single spiral arm is distinguished. The density along this stream is very low, because for simulation V150a5 the gas density is much lower than for the other simulations and only a small fraction of the stellar wind is focused into the accretion wake. Also notice that given the
low density of the gas, the effective resolution of this model is lower.  Given the large velocity of the wind compared to the orbital velocity of the system, the wind escapes the system without being affected by the presence of the companion star and remains spherically symmetric. 

If we compare simulations V15a5 and V30a5 we see that the angle formed by the inner high-density spiral arm and the axis of the binary changes as a function of the 
velocity of the wind. For the low-velocity model, this angle is very sharp and as the wind velocity increases the angle of the stream becomes more oblique. 
This is expected given that the angle of the accretion wake $\theta$ depends on the apparent wind direction seen by the accretor in its orbit, such that $\tan{\theta} = v_\mathrm{orb}/v_\mathrm{wind}$.

\subsubsection{Accretion disk}\label{sec:accretion_disk}

\begin{figure*}
\centering
\includegraphics[width=0.9\textwidth]{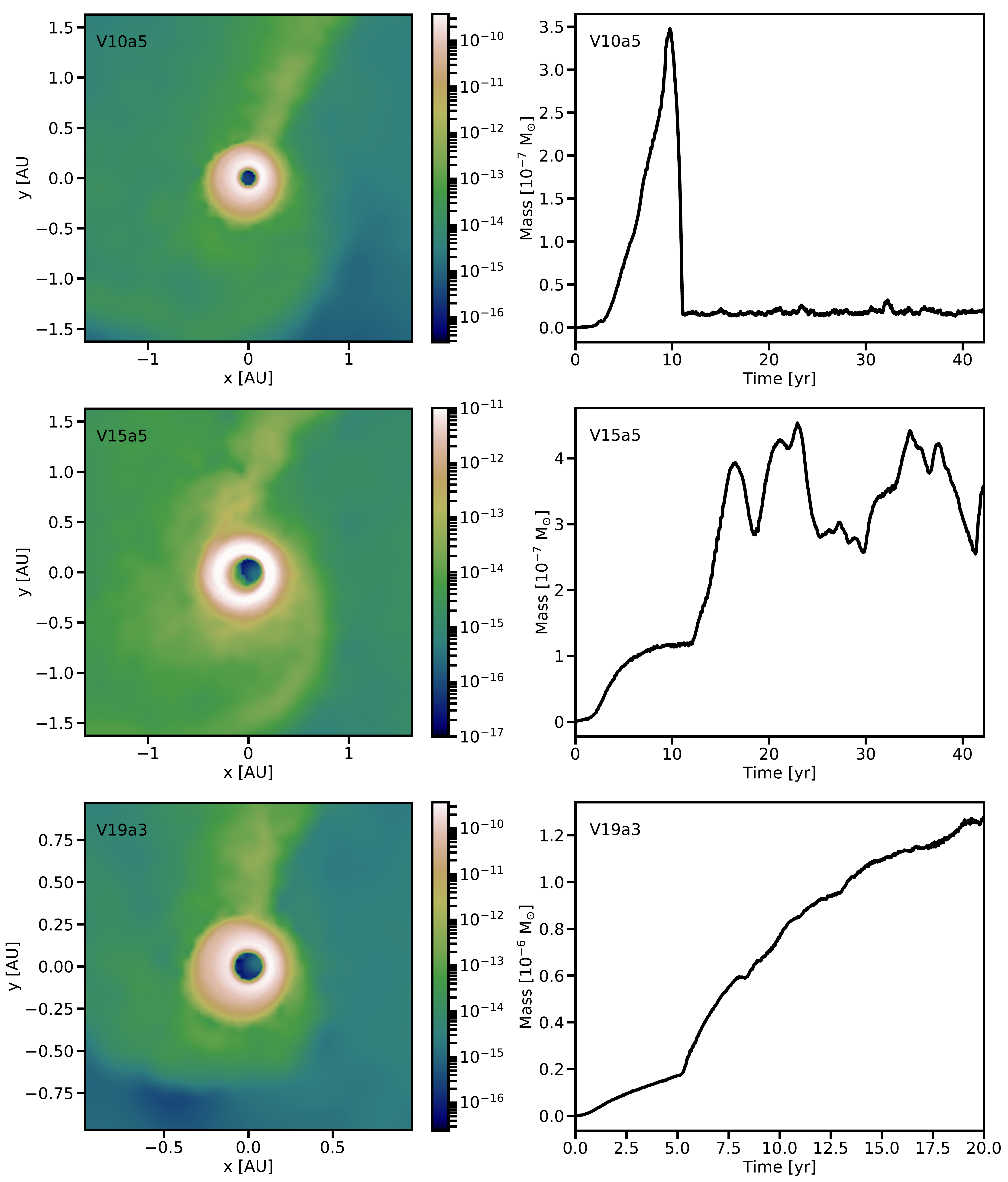}
\caption{The left-hand plots show the density in units of g cm$^{-3}$ in the orbital plane of the region centered on the accreting star for three models with  $v_{\infty}/v_\mathrm{orb} < 1$, zooming in on the accretion disks. The size of the box corresponds to the Roche-lobe diameter of the companion star. For models V15a5 and V19a3, the situation after 7.5 orbits is shown, whereas for model V10a5 we show the accretion disk before it disappears, but in the same orbital phase. In the right-hand panels we show the disk mass, i.e. the mass contained in a sphere of radius $R_\mathrm{disk}$ (see Table \ref{table:main}) as a function of time.}
\label{fig:disks}
\end{figure*}

An accretion disk is formed around the accretor in the simulations with $v_{\infty}/v_\mathrm{orb} < 1$ (V10a5, V10a5s2 V15a5, V15as2 and V19a3). 
Figure \ref{fig:disks} shows the density of the gas in the orbital plane for the accretion disks formed around the secondaries for simulations V10a5, 
V15a5 and V19a3. 
One common feature is that the accretion disks are not symmetric. This asymmetry is associated with wind coming from the donor star colliding with material already present in the disk. Another feature in the flow pattern of the simulations is formed by two streams of gas feeding the disk: as seen from the velocity field in the corotating frame (not shown), one stream arises from material moving along the inner spiral arm with $v_{x}, v_{y} < 0$ and the second one comes from material moving along the outer spiral arm with $v_{x}, v_{y} > 0$. 

Even though the accretion disk is not symmetric, we approximate it as such in order to estimate its size and mass. We construct spheres of different radii centered at the position of the companion star and measure the mass within each sphere as a function of time. The size of the disk is taken as the radius of that sphere outside which the mass is approximately constant as a function of radius. The disk mass is taken as the total gas mass within that sphere. This approach is reasonable because the gas mass inside the 
spheres is dominated by the mass of the disk. Table \ref{table:main} shows the resulting values of the accretion disk radius and mass for the simulations in which a disk was present after eight orbital periods, except for model V10a5 where the given radius and mass correspond to a time of 2 orbits.   

In the right-hand panels of Figure \ref{fig:disks}, we show the mass within the disk as a function of time for the same models. For model V15a5 (middle right), it is clear that the disk mass is variable over time. It is not obvious where this variability comes from, but we note that the outer disk radius is not much larger than the sink radius (34.7 R$_{\odot}$) which determines the inner edge. To investigate this behaviour we performed a simulation with the same parameters but a smaller sink radius (20.8 R$_{\odot}$, V15a5s2). In this simulation the disk reaches a larger radius and gradually grows in mass, up to a value at the end of the simulation about 9 times larger than in V15a5 (see Table \ref{table:main}). This can be understood as the smaller sink allows gas in the disk to spiral in to closer orbits around the secondary by viscous effects, transferring angular momentum to the outer regions of the disk which thus grows in size. Such behaviour is physically expected, but simulations with even smaller sink radii and extending over a longer time are required to investigate the mass and radius of the disk and their possible variability.

For model V10a5 (top right panel of Fig. \ref{fig:disks}), we observe a highly dynamic accretion disk, which disappears after two orbital periods. The moment when the disk vanishes can be seen as a steep decrease in the mass of the accretion disk. The reason for this disappearance is that after two orbital periods the size of the outer radius of the disk becomes smaller than the radius of the sink particle. When performing a similar simulation but with a smaller sink radius,
V10a5s2, the accretion disk remains for the rest of the simulation and gradually increases in mass, similar to what was seen in simulation V15a5a2 discussed above. The size of the disk remains smaller than in the models with a wind velocity of 15 km s$^{-1}$. A similar pattern is found in model V19a3 (bottom
panel of Figure \ref{fig:disks}), in which the disk mass increases with time. A longer simulation will be needed to see if it converges towards a constant value.

For the models in which $v_{\infty}/v_\mathrm{orb} > 1$, no accretion disk is formed around the companion.
Also in a test simulation with a smaller sink particle (V30a5s2) the accreted matter is swallowed by the sink particle without forming an accretion disk. We conclude that if an accretion disk were to form in this system it will be smaller than the assumed sink radius. 
 
\begin{figure*}
\centering
\begin{minipage}[b]{.45\textwidth}
\includegraphics[width=\hsize]{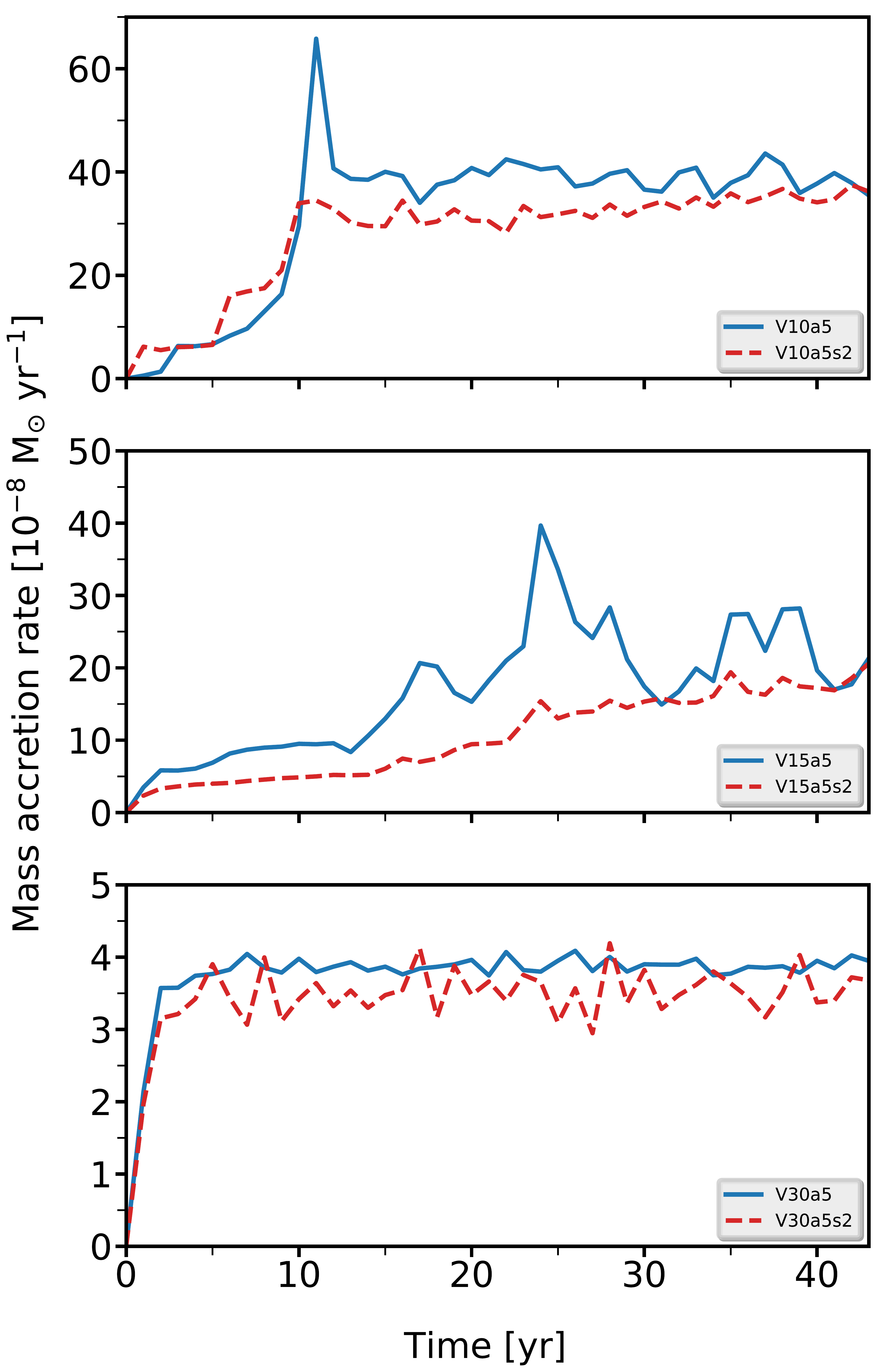}
\caption{The average accretion rate over one-year intervals for models V15a5 (top), V10a5 (middle) and V30a5 (bottom) is shown in blue. The dashed red lines correspond to the year-averaged mass accretion rate for the corresponding models with a smaller sink radius.}\label{fig:mass_accretion}
\end{minipage}\qquad
\begin{minipage}[b]{.45\textwidth}
\includegraphics[width=\hsize]{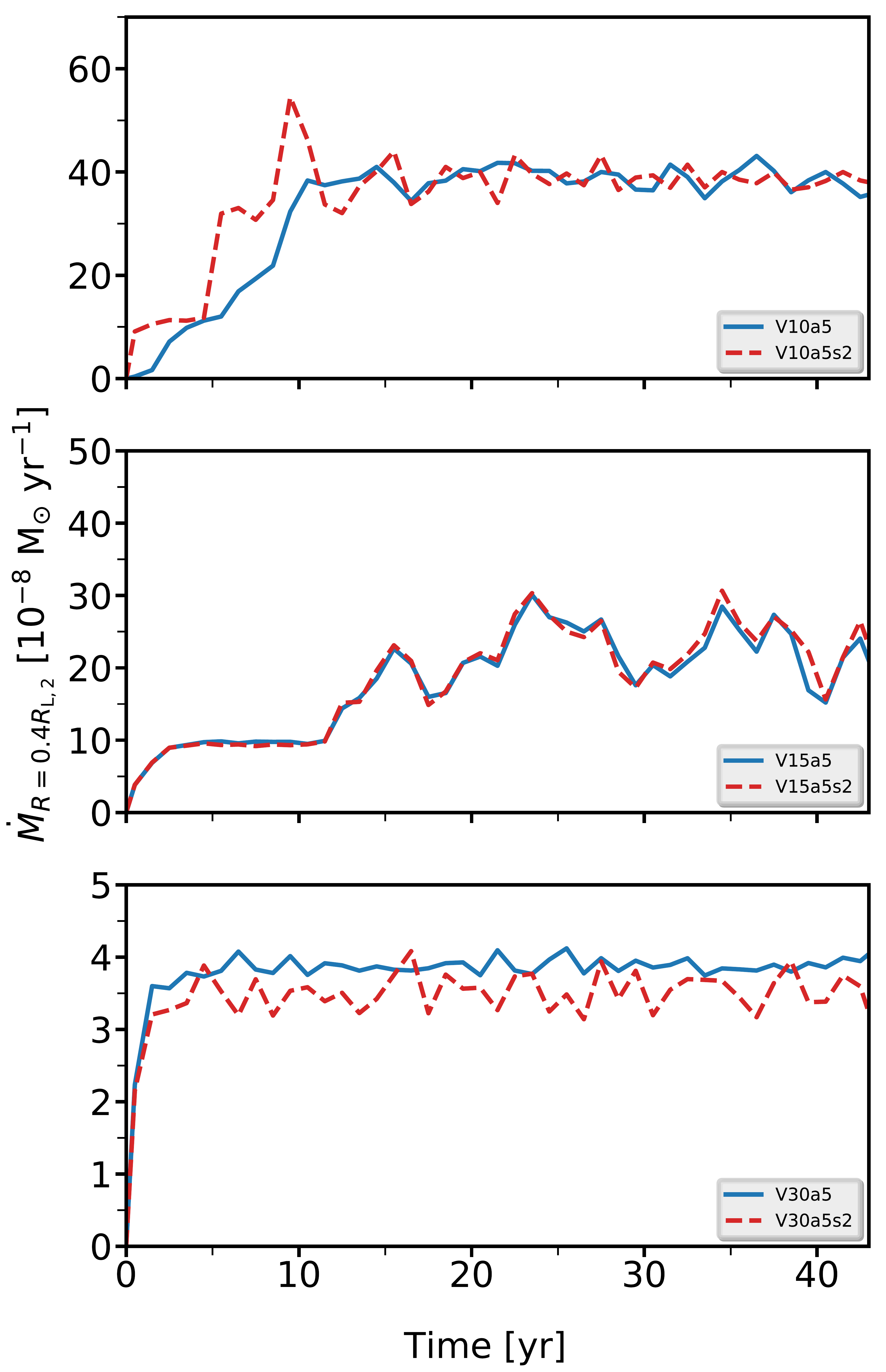}
\caption{The net inflow rate into a sphere of radius 0.4$R_\mathrm{L, 2}$ around the accretor star averaged over intervals of one year for the same models as in Figure \ref{fig:mass_accretion}.\vspace{0.8cm}}\label{fig:mass_flux}
\end{minipage}
\end{figure*}

\subsubsection{Mass-accretion rate}\label{sec:mass_accretion_rate}

Figure \ref{fig:mass_accretion} shows the average mass-accretion rate over intervals of one year for models V10a5, V15a5, V30a5 (solid lines), V10a5s2, V15a5s2 and V30a5s2 (dashed lines). 

In models V15a5 (middle panel) and V19a3 (not shown), we find a lot of variation in the accretion rate, which clearly exceeds the noise associated with the finite number of particles accreted. This suggests a correlation between the mass of the accretion disk and the accretion rate. This can be observed if we compare the middle right panel of Figure \ref{fig:disks} and the middle panel of Figure \ref{fig:mass_accretion}, where the peaks and troughs in the mass of the disk coincide in time with those observed in the mass-accretion rate. In the case of model V15a5s2, where the mass in the disk gradually increases with time, the accretion rate as a function of time also shows an increase (dashed line in the middle panel of Fig. \ref{fig:mass_accretion}). The average accretion rate for model V15a5s2 is a factor of 1.5 lower than for model V15a5 where the
sink radius is larger. For model V10a5 (top panel of Figure \ref{fig:mass_accretion}), we find that after the accretion disk becomes smaller than the sink and is swallowed, the variation in the accretion rate decreases. For model V10a5s2 with a smaller sink, although the accretion disk is present for the entire simulation, the variability in the accretion rate is smaller than for model V15a5s2 (where the accretion disk is also present over the entire simulation). However, we should note that the resolution for model V10a5s2 is much lower. Similar to models V15a5 and V15a5s2, the mass-accretion efficiency is lower for model V10a5s2 than for model V10a5.
In the bottom panel of the same figure, we show the accretion rate for model V30a5. Apart from the Poisson noise associated with the finite resolution of the SPH simulations, the accretion rate is constant in time. For the same simulation but with a smaller sink we find a slightly lower accretion efficiency. 

We find that the mass that is not accreted in models V10a5s2 and V15a5s2 is stored in the accretion disk. Figure \ref{fig:mass_flux} shows the net mass inflow rate into a shell of radius $0.4 \mathrm{R}_\mathrm{L,2}$ around the secondary for models V10a5, V15a5 and V30a5 (solid lines) and their corresponding models with a smaller sink radius
(dashed lines). The chosen shell radius is larger than the size of the accretion disk in all the simulations, so that the inflow rate corresponds
to the total mass-accumulation rate of the sink and the disk, $\dot{M}_{0.4\mathrm{R}_\mathrm{L}} = \dot{M}_\mathrm{sink} + \dot{M}_\mathrm{disk}$, since the amount of gas within this volume that is \emph{not} in the disk is very small. In the steady state, the inflow rate we find in this way is insensitive to the chosen sink radius. 
For simulations V15a5 and V15a5s2, this combined accretion rate is seen to be very similar and also shows similar variability over time. A similar feature holds for simulation V10a5 and V10a5s2. Since it is likely that the mass in the disk will eventually 
reach the stellar surface, this may be a better measure of the steady-state accretion rate that is insensitive to the assumed sink radius. In the case of models V30a5, V30a5s2, we observe the same small  discrepancy in the inflow rate at $R = 0.4$R$_\mathrm{L,2}$ that we find in Figure \ref{fig:mass_accretion}. The net influx of mass within the shell is the same in both simulations, reflecting the steady-state condition reached. However, when the sink radius 
is larger it captures material that may otherwise escape. Therefore, for simulations such as V30a5 in which no accretion disk is formed, our results provide only an upper limit to the amount of mass accreted.  

\begin{figure}
\centering
\includegraphics[width=\hsize]{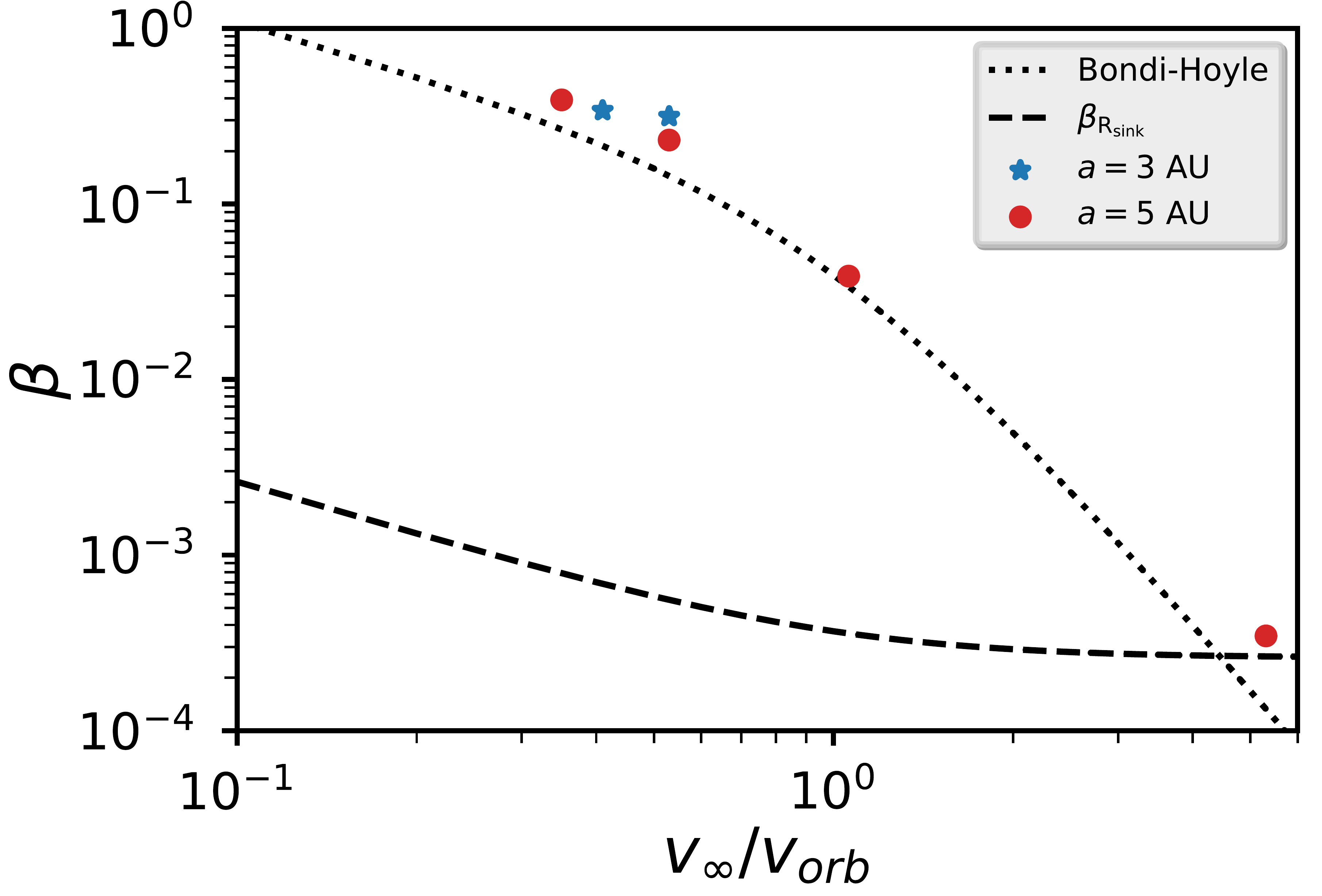}
\caption{The fraction of mass accreted by the companion as a function of $v_{\infty}/v_\mathrm{orb}$. The dotted line corresponds to BHL accretion rate with $\alpha_\mathrm{BHL} = 1$, and the dashed line to the accretion rate for a geometrical cross section of size $R_\mathrm{sink}$. Dots correspond to models in which the orbital separation is 5 AU and stars to models in which $a = 3$ AU. The values shown are measured at a radius of 0.4 $R_\mathrm{L,2}$ ($\beta_{0.4 R_\mathrm{L,2}}$).}
\label{fig:beta}
\end{figure}

In order to determine the long term evolution of the orbit, one of the quantities of interest from these simulations is the average mass-transfer efficiency $\beta$. 
Given the results discussed above and shown in Figures \ref{fig:mass_accretion} and \ref{fig:mass_flux}, we take the mass-transfer efficiency as $\beta_{0.4\mathrm{R}_\mathrm{L}} = \dot{M}_{0.4\mathrm{R}_\mathrm{L}}/\dot{M}_\mathrm{d}$. In Table \ref{table:main}, we provide the mean values of this quantity for the science simulations, as well as the mean values for the material captured by the sink only, i.e. $\beta_\mathrm{sink} = \dot{M}_\mathrm{sink}/\dot{M}_\mathrm{d}$.
In Figure \ref{fig:beta}, we show the corresponding values for the accretion efficiency $\beta_{0.4\mathrm{R}_\mathrm{L}}$ as a function of the ratio $v_{\infty}/v_\mathrm{orb}$ for the science simulations.  The dotted line corresponds to $\beta_\mathrm{BHL}$ in equation \ref{BHL} with $\alpha_\mathrm{BHL} = 1$. Solid dots in the figure correspond to models V15a5-V150a5 and stars to models T3 and V19a3, both of which have $a = 3$ AU. 

Model V150a5, in which we use a wind velocity ten times larger than the typical velocities of AGB winds, was set up to approximate the isotropic wind mode. Figure \ref{fig:beta} shows that for this model, our numerical result  for the accretion efficiency exceeds the expected value for BHL accretion by a factor of 2.5 (see also Table \ref{table:main}). However, we note that the BHL accretion radius for this simulation is smaller than the radius of the sink, which will result in a discrepancy with the BHL prediction for the accretion efficiency. The dashed line in the same figure corresponds to the accretion efficiency $\beta_\mathrm{sink}$ assuming the geometrical cross section of the sink instead of the BHL cross section. The difference between the simulation result and $\beta_\mathrm{sink}$ is reduced to a factor of 1.3. 

Fig. \ref{fig:beta} shows several interesting features. 
As the ratio $v_{\infty}/v_\mathrm{orb}$ increases, the value of $\beta$ decreases, following a similar trend as expected from BHL accretion. However, for $v_{\infty}/v_\mathrm{orb} < 1$ we find $\beta$ values that exceed the $\beta_\mathrm{BHL}$ by up to a factor of 1.5-2.3. For the lowest wind velocity, the mass transfer efficiency is approximately 40\%. 
Finally, for models V15a5 and V19a3, which have the same $v_{\infty}/v_\mathrm{orb}$ ratio, the accretion efficiencies shown in Figure \ref{fig:mass_flux} differ by a factor of 1.4.

\subsubsection{Angular-momentum loss}

\begin{figure}
\centering
\includegraphics[width=\hsize]{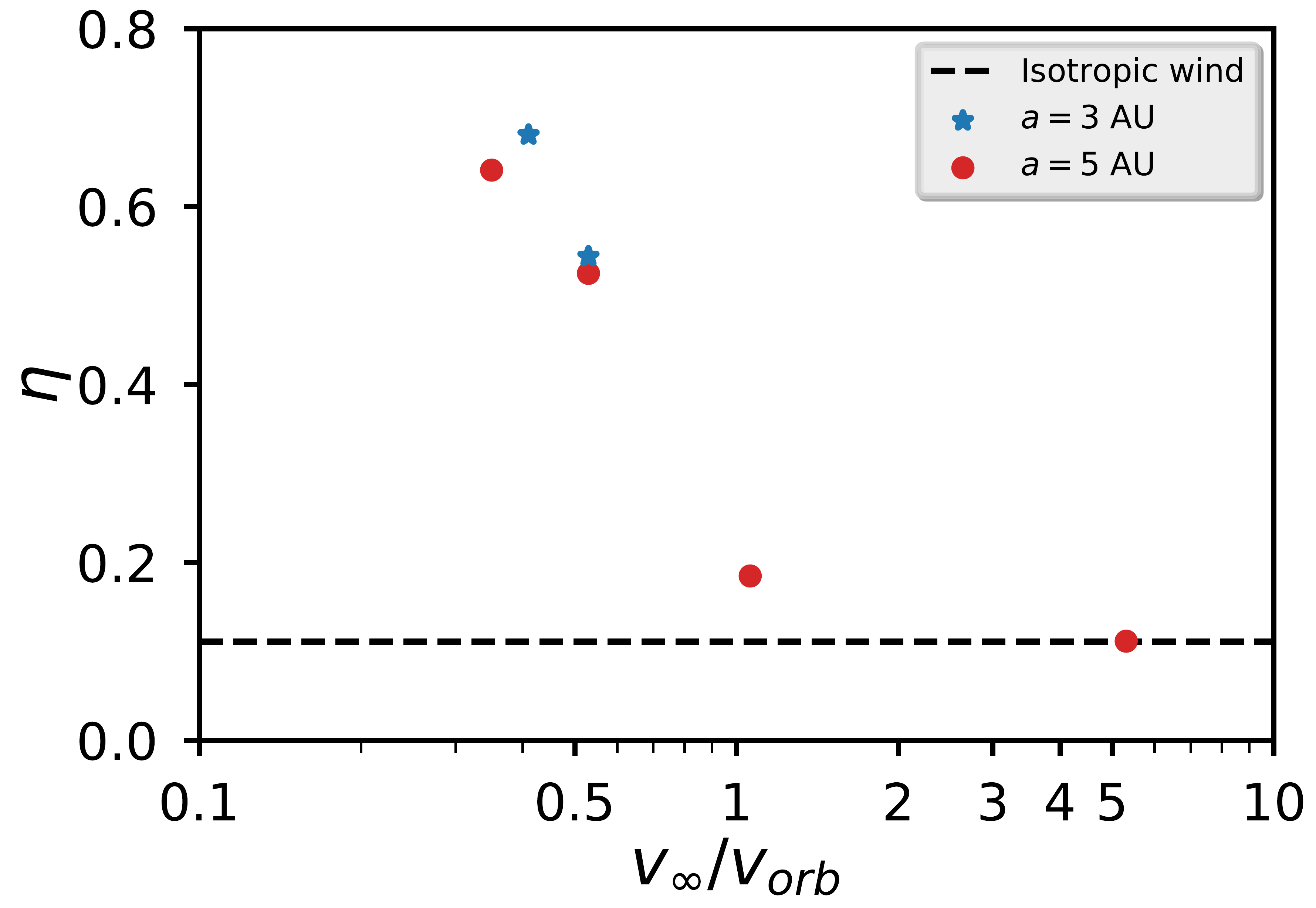}
\caption{The specific angular momentum of the mass lost from the system as a function of $v_{\infty}/v_\mathrm{orb}$. The smaller the velocity of the wind compared to the orbital velocity, the higher the specific angular momentum loss. For equal velocity ratios, the specific angular momentum is very similar. The dashed line shows the value $\eta_\mathrm{iso}$ expected for an isotropic wind for a mass ratio $q=2$.}
\label{fig:eta}
\end{figure}

The second quantity of interest resulting from our simulations is the value of the specific angular momentum of the material lost by the system. Figure \ref{fig:eta} shows the values of this quantity in units of 
$J/\mu$ as a function of $v_{\infty}/v_\mathrm{orb}$ for the different models. We see that the smaller the velocity of the wind, the larger the specific angular momentum of the material lost. 
This is consistent with the expectation that when the velocity of the wind is small compared to the relative orbital velocity of the system, more interaction between the gas and the stars occurs and more
angular momentum can be transferred to the gas. 
This strong interaction is also seen in the large mass-accretion
efficiency, as well as in the existence of a large accretion disk and in the structure of the spiral arms of the systems.
The simulations with $v_{\infty} < v_\mathrm{orb}$ show spiral arms that are more tightly wound around the system, and with higher gas densities, than the simulations for higher wind velocities (see Sect. \ref{morphology} and Fig. \ref{fig:overall}). These spiral arms correspond to the accretion wake of gas interacting with the secondary star. We interpret the loss of angular momentum as the result of a torque between the gas in this accretion wake and the binary system. The magnitude of this torque depends on the orientation of the wake and the density of the gas behind the companion star. When the accretion wake is approximately aligned with the binary axis and the density in the wake is relatively low, as in simulations with $v_{\infty}/v_\mathrm{orb}>1$,  the torque between the gas and the binary will be small. The torque exerted on the gas will increase as the wake misaligns with the binary axis, and as the gas density in the accretion wake becomes higher. Both these effects occur for low values of $v_{\infty}/v_\mathrm{orb}$, in particular in the inner spiral arm that has a high density and is oriented almost perpendicular to the binary axis (Fig. \ref{fig:overall}).Therefore, the transfer of angular momentum from the orbit to the outflowing gas increases strongly with decreasing 
$v_{\infty}/v_\mathrm{orb}$.Furthermore, it implies that most of the angular-momentum transfer occurs at short distances from the binary system, as is confirmed by the test discussed in Section \ref{sec:steady} (see Figure \ref{fig:mass_loss}).

Model V150a5 approximates the isotropic or fast wind mode fairly well. In this case, the ejected matter escapes from the system with very little interaction, only removing the specific angular momentum of the orbit of the donor star. 
For this simulation, the value we obtained for $\eta$ is $\eta_\mathrm{num}=0.112\pm0.001$, very close to the expected value for the mass ratio of the stars in our models, $\eta_\mathrm{iso} =  0.111$ (dotted line in Fig. \ref{fig:eta}). 

An interesting result of our models is that for the same value of $v_{\infty}/v_\mathrm{orb}$, but different orbital separations, the specific angular momentum of the mass lost is very similar. For models V15a5 $\eta \approx 0.52$ and for V19a3, $\eta \approx 0.54$. Model V10a5 has the largest loss of angular momentum among the high-resolution simulations with $\eta = 0.64$, although the low-resolution simulation T3 has an even large amount of angular-momentum loss, $\eta = 0.68$. 

\section{Discussion}\label{sec:discussion}

\subsection{Assumptions and simplifications}

A complete study of binary systems interacting via AGB winds requires the simultaneous modelling of hydrodynamics, radiative transfer, dust formation and gravitational dynamics. 
In this study, several simplifications have been made. 

\subsubsection{Wind acceleration}\label{sec:wind_acceleration}

Dust formation and radiative transfer are not included in our models (except for the simplified effective cooling model) although these processes play a major role in accelerating the gas away from the star. 
Instead of explicitly computing the gas acceleration, a constant velocity of the gas is assumed with values typical of the terminal velocities of AGB winds. This method is chosen to simplify the calculations and to guarantee that the velocity of the wind $v_{\infty}$ at the location of the companion star has a predefined value as we want to study the effect of the velocity of the wind on the interaction of the gas with the stars. 
On the other hand, observations and detailed wind models both indicate that most of the acceleration of AGB winds occurs in the dust-formation zone located at $R_\mathrm{dust} \sim 2-3 R_{\star}$ \citep{Hoefner+2018}. For the orbital parameters assumed in our models, $R_\mathrm{dust}$ is similar to or larger than the Roche-lobe radius of the donor star, $R_\mathrm{L,1}$. As described in \citet{shazrene_thesis} and \citet{carlo1}, in this situation mass transfer can occur by WRLOF, i.e. by a dense flow through the inner Lagrangian point into the Roche lobe of the companion. However, given that the wind particles in our simulations are injected with predefined terminal velocities from the radius of the donor star, no WRLOF is observed. 

If the gradual acceleration of the wind is taken into account, the density and velocity structure of the gas inside the orbit of the companion will be different which may affect the outflow morphology in our low wind-velocity simulations.
To some extent the effects of wind acceleration can be accounted for by assuming an increasing velocity profile in Eq. \ref{numerical1}, mimicking the velocity profile of an AGB wind. The resulting acceleration in such models would still be spherically symmetric. In addition, the modified outflow in a binary can change the optical depth of the dust and thus affect the acceleration of the wind itself, making it non-isotropic. This can only be studied by explicitly modelling dust formation and radiative transfer in wind mass-transfer simulations \citep[e.g. see][]{shazrene,Chen+2017} and is beyond the scope of the present study.
For this reason, and although the simulations presented here give insight into the dependence of the angular-momentum loss and the accretion efficiency on the $v_{\infty}/v_\mathrm{orb}$ ratio, our results should be interpreted with some caution. 

\subsubsection{Accretion}
\label{sec:accretion}

Given the small radius of a main-sequence or white-dwarf companion star, resolving it spatially would be very computationally demanding. 
For this reason the accretor was modelled as a sink particle with radius equal to a fraction of the size of its Roche lobe. 
In the simulations where an accretion disk is formed, the inner disk radius is limited by the sink radius. However, physically there is no reason to expect that the disk will not extend inwards to the stellar radius by viscous spreading, and any gas captured in the disk will be transferred towards the stellar surface. As we have shown in Section \ref{sec:mass_accretion_rate}, the sum of the accretion rate onto the sink and the mass accumulation rate in the disk appears to be independent of the assumed sink radius, and should give a reliable measure of the mass transfer rate to the companion. 
On the other hand, we cannot be sure that all matter that approaches the surface of the star through the disk will be accreted, as some fraction of it may be expelled by a wind or jet formed in the inner part of the accretion disk. 
In simulations with larger wind velocities where we do not find an accretion disk, it is possible that a disk is still formed at a smaller radius within the sink. However, some of the mass captured within the sink may not end up in the disk and reach the surface of the companion.
For these reasons, and although our results for the mass accretion efficiency with different prescriptions for the artificial viscosity and sizes of the sink particle seem to converge, the values provided should be taken as upper limits. 

\subsubsection{Angular-momentum transfer}
\label{sec:angular_momentum_disc}

In our simulations the gravitational field of the gas is neglected, i.e. the gas feels the gravitational potential of the stars but the stars do not feel the gravity of the gas. This does not prevent us from inferring the gravitational influence of the gas on the binary system, because conservation of angular momentum dictates that the angular momentum transferred to the gas, which we measure from the simulation, is taken out of the orbit. This is expressed by Equation \ref{eq:a} in Sect. \ref{sec:orbit}. We have performed test simulations in which the gravitational forces are symmetric, i.e. the stars do feel the gravity of the stars, and find that the results for the gas dynamics and angular-momentum loss are not affected. We also find (see Sect. \ref{sec:convergence}) that the specific angular momentum of the outflowing gas is very well conserved once it travels beyond a few orbital radii. This is an advantage of using an SPH code for this work, because SPH codes are much better at conserving angular momentum than grid-based hydrodynamical codes. Therefore, we are confident that our results for the angular-momentum loss are robust and insensitive to the numerical assumptions in the simulations.

\subsubsection{Rotation}
\label{sec:rotation}

The stars in our simulations are assumed to be non-rotating. 
As mentioned in Section \ref{sec:method}, given the evolutionary history of expansion and mass loss of the AGB star, we expect the rotation rate of the donor to be negligible compared to the orbital frequency, as long as tidal interaction 
can be ignored. However, the tidal synchronization timescale of the donor can be fairly short. Applying the equilibrium tide model as described in \citet{hurley} and using the orbital and stellar parameters of the simulated systems (in
particular, a stellar radius of 200 R$_{\odot}$) we find a synchronization timescale of several times $10^4$ years for an orbital separation of 5\,AU, and about $10^3$ years for 3\,AU. This is much shorter than the expansion timescale of the
AGB star, which is of the order of $10^6$ years, indicating that the donor is likely to rotate synchronously with the orbit. As a consequence, an additional transfer of angular momentum to the gas will take place at the expense of the 
rotational angular momentum of the donor. This occurs at a rate:
\begin{equation}
\label{eq:Jdot_spin}
\dot{J}_\mathrm{rot} = \frac{2}{3} R_\mathrm{d}^2\Omega\dot{M}_\mathrm{d},
\end{equation}
if we assume the wind decouples from the star at a spherical shell of radius $R_\mathrm{d}$. In the absence of magnetic or hydrodynamic coupling between the outflowing wind and the star, $R_\mathrm{d}$ can be identified with the stellar radius. If the donor remains tidally locked to the orbit, this angular momentum is continually supplied to the gas from the orbit, leading to additional orbital angular-momentum loss compared to the non-rotating case. We choose not to include this in our simulations, because the effects of spin angular-momentum loss and tidal interaction can be taken into account separately, as is often done in binary population synthesis modelling \citep[e.g.][]{hurley,carlo4}. For the systems we have simulated, the rotational angular-momentum loss implied by Eq.~\ref{eq:Jdot_spin} is only a small fraction of the orbital angular-momentum loss, so the effect on the orbit will be small. By ignoring rotation, however, we also neglect the possibility that the morphology of the outflow itself is modified, and thereby the way it interacts with the companion. However, if the wind decouples from the donor star at a radius much smaller than the orbital separation, the flow in the vicinity of the companion will not be strongly affected.

On the other hand, the companion star will gain not only mass but also angular momentum from the material accreted which will lead to spin-up of the star. Once the star is spun up to its critical rotation, no more accretion can take place \citep{packet, matrozis}, imposing a constraint on the amount of accreted material. 
In our simulations we keep track of the angular momentum accreted by the sink particle. However, this is not representative of the true angular momentum added to the companion because we use a sink radius that is much larger than the expected stellar radius. In cases where an accretion disk is formed, the angular momentum of the accreted gas corresponds to a Kepler orbit at the sink radius, whereas in reality by the time the gas reaches the surface of the star it would have transferred angular momentum to the outer regions of the disk. Therefore our simulations do not allow us to study the effects of angular-momentum accretion on the secondary \citep{Liu+2017}.

\subsection{Comparison to other work}\label{sec:comparison}

\subsubsection{Angular-momentum loss}

Our results for the angular-momentum loss are in approximate agreement with other work which uses different methods. 
\citet{jahanara2} performed 3D hydrodynamical grid simulations of a star undergoing mass loss and interacting with a companion star. 
They study the amount of angular-momentum loss as a function of the wind speed at the surface of the donor for various mass ratios and different assumed mass-loss mechanisms. 
The mechanism that best approximates the mass loss from an AGB star is the radiatively driven (RD) wind mechanism. This is roughly equivalent to our assumptions, although \citet{jahanara2} use an adiabatic EoS without cooling and they assume a much higher sound speed than in our simulations, which leads to substantial gas-pressure acceleration in their models at low wind velocities. 
Despite these differences, and even though the mass ratio in our simulations, $q=2$, is different from the mass ratios assumed in \citet{jahanara2}, we can make a rough comparison of our results with their Figure~7, in which $q =1$. 
In that figure $V_\mathrm{R} = v_\mathrm{w,RL}/v_\mathrm{orb}$ corresponds to the average wind velocity at the Roche-lobe surface of the donor, which can be compared to our $v_{\infty}/v_\mathrm{orb}$ ratio. The parameter $\ell_w$ corresponds to the specific angular momentum of the material lost in units of $J/\mu$ and is equivalent to our $\eta$, although in \citet{jahanara2} the donor is kept in corotation with the orbit, so that $\ell_w$ includes the spin angular
momentum loss from the donor star. 
We compare these results to our simulations in Figure \ref{fig:eta_fit}, where we have subtracted the small amount of spin angular momentum-loss from the $\ell_w$ values using Equation \ref{eq:Jdot_spin}, so that they only represent the orbital angular momentum loss.
\citet{jahanara2} find that the strongest angular-momentum loss, with $\ell_w \approx 0.6$, occurs for the lowest wind velocities, corresponding to $V_\mathrm{R} \approx 0.7$, which is comparable but slightly larger than our results (see Figure \ref{fig:eta_fit}).  For increasing wind velocity the specific angular momentum decreases, converging to values equal to the isotropic case for $q=1$, i.e. $\eta =0.25$, which is consistent with our results for model V150a5.

\citet{Chen+2017,rochester} also performed grid code simulations of binary systems interacting via AGB wind mass transfer in order to study the orbital evolution, but including pulsations of the AGB star as well as cooling, dust formation and radiative transfer. 
They modelled systems consisting of a primary star of 1~$M_{\odot}$ with a terminal wind velocity $v_{\infty} \approx 15$ km s$^{-1}$ and orbital separations between 3 and 10~AU. Most of their models have a secondary mass of 0.5~$M_{\odot}$, i.e. the same mass ratio as in our simulations. 
Their models with $a > 6$~AU display a similar morphology to our simulations with $v_{\infty}/v_\mathrm{orb} \ga 1$, showing a spiral arm structure corresponding to the BHL accretion wake of the secondary. On the other hand, their models with smaller separation (and $v_{\infty}/v_\mathrm{orb} \la 1$) show a flow morphology resembling WRLOF and appear to be forming a circumbinary disk, which we do not find in our low-velocity simulations. These differences are likely due to the differences in modelling the wind acceleration process.
\citet{rochester} express the angular momentum loss from the system in terms of a parameter $\gamma$, which is the specific angular momentum of the matter lost in units of $J/M_\mathrm{bin}$. This is equivalent to our description in terms of $\eta$, using the transformation
\begin{equation} \label{eq:eta-gamma}
\eta = \frac{\mu}{M_\mathrm{bin}}\,\gamma = \frac{q}{(1+q)^2}\, \gamma.
\end{equation}
They find larger angular momentum loss from systems with smaller $a$, i.e. with smaller $v_{\infty}/v_\mathrm{orb}$, similar to our results. For their models with $q=2$ and $a \geq 6$~AU (which have $1.0 \la v_{\infty}/v_\mathrm{orb} \la 1.3$) they find similar values of $\eta$ to our model V30a5 with $v_{\infty}/v_\mathrm{orb} = 1.06$. However, for the model with the same mass ratio and $a = 4$~AU ($v_{\infty}/v_\mathrm{orb} \approx 0.8$, intermediate between our models V15a5 and V30a5), they obtain a higher $\eta \approx 0.6$ than we find for V15a5 which has lower $v_{\infty}/v_\mathrm{orb}$.\footnote{Interestingly, their 3~AU model with $q=10$ has almost the same $v_{\infty}/v_\mathrm{orb}$ and $\eta$ value as the 4~AU model with $q=2$, although the equivalent $\gamma$ value is different.}
In making the comparison with our models we have to take into account that in the \citet{rochester} simulations the AGB star spin is synchronised with the orbit. 
The spin angular momentum transferred to the gas in their simulations is substantially larger than given by Eq.~\ref{eq:Jdot_spin}, which they ascribe to subsonic turbulence between the photosphere and the dust formation radius. When we subtract the contribution of the stellar spin (as given in their Table~4) from the total angular-momentum loss, the resulting $\eta$ values (in particular for $a=4$~AU) are in better quantitative agreement with our results (see open squares in Figure \ref{fig:eta_fit}).

\subsubsection{Accretion efficiency}

Several previous studies have investigated the mass-accretion efficiency during wind mass transfer. \citet{theuns2} and \citet{Liu+2017} performed SPH simulations of a binary with exactly the same parameters as our test models (T1--T3), using an EoS without cooling. They find accretion efficiencies between 1\% and 2.3\% when using an adiabatic EoS, which is quite comparable to the low $\beta$ value of our model T1 (and substantially less than expected from BHL accretion, see also \citealp{Nagae+2004}). On the other hand, when they apply an isothermal EoS the accretion efficiency increases to 8\% and 11\%, respectively. This is about a factor of three smaller than we find in our isothermal test model T2 and in model T3 that includes gas cooling explicitly. The reason for this difference is unclear, but we note that the algorithm for computing the accretion rate in these papers is different than the method we use in this work. Besides the discrepancies in the accretion efficiencies, these studies and ours both indicate that allowing for gas cooling results in substantially larger accretion efficiencies.  

As discussed in Section \ref{sec:wind_acceleration}, our simulations lie in a regime where WRLOF might be expected if wind acceleration were taken into account. In the WRLOF regime accretion onto the companion star occurs by a combination of material flowing through the inner Lagrangian point and gravitational focusing towards the companion star. Our simulations show the latter effect, but not the former.
In the WRLOF simulations by \citet{shazrene_thesis} (see also \citealp{carlo1}) and \citet{Chen+2017}, which also include gas cooling, large accretion efficiencies of up to 40--50\% are found. It is interesting that we find quite similar $\beta$ values in our lowest wind-velocity models, even though there is no significant flow though the inner Lagrangian point in our models. This suggests that the high accretion efficiencies found in WRLOF simulations may be caused predominantly by gravitational focusing onto the companion star.
However, given the uncertainties in reliably determining accretion rates as discussed in Sects. \ref{sec:mass_accretion_rate} and \ref{sec:accretion}, one should be cautious in making such comparisons.

\subsection{Implications for binary evolution}

\begin{figure}
\centering
\includegraphics[width=\hsize]{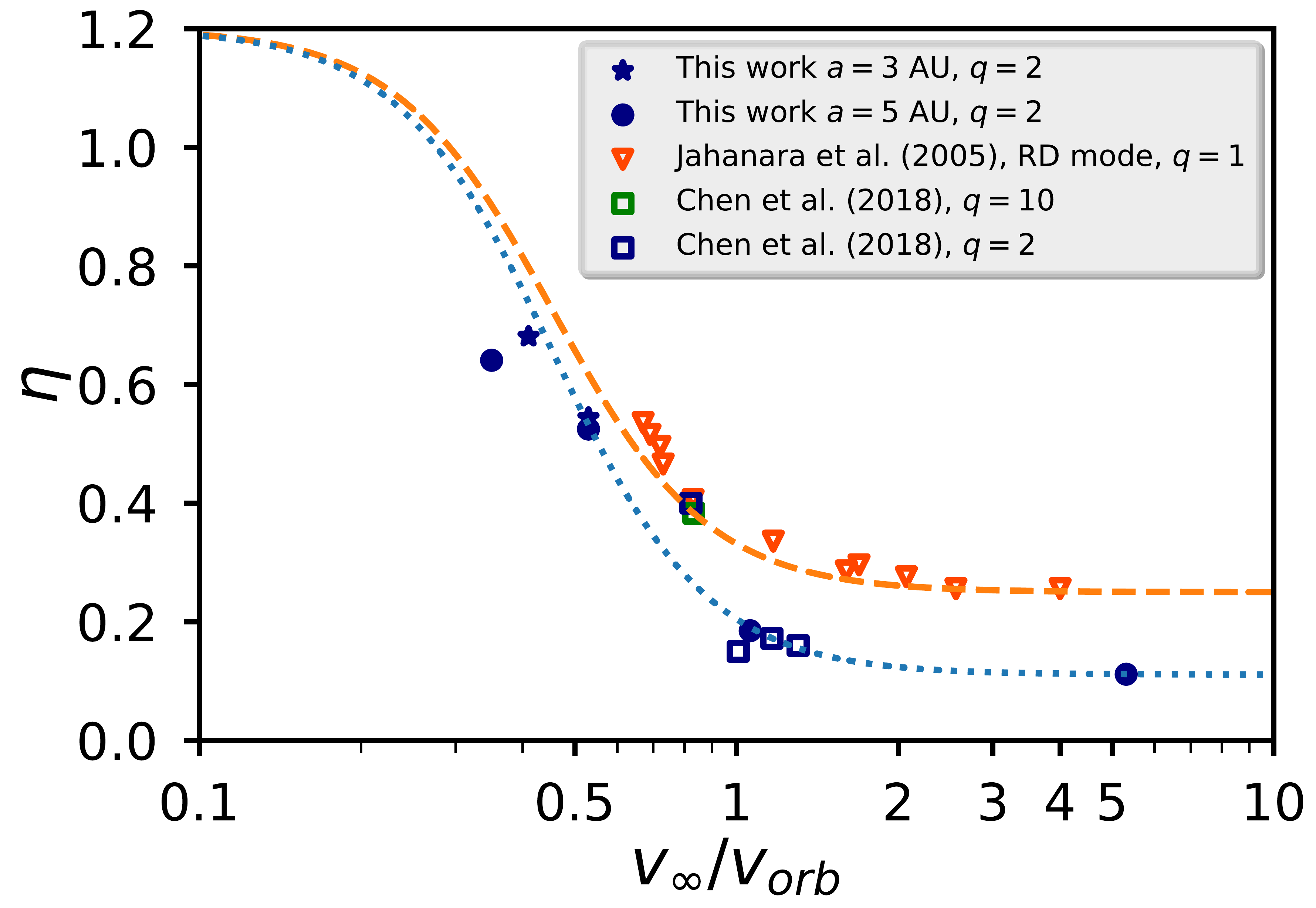}
\caption{Comparison of our results (Fig. \ref{fig:eta}) to other work. The data points correspond to the specific orbital angular momentum lost from the system via winds obtained using different methods. The open squares correspond to the results of \citet{rochester}, triangles show the results of \citet{jahanara2} for the radiatively driven wind mode. In the latter case we have interpreted the wind velocity at the Roche-lobe surface as corresponding to $v_\infty$. The results from both papers have been corrected for the spin angular-momentum loss (see text) to compare with our results. The colors correspond to different mass ratios $q$ and the lines to the fit formula described by equation \ref{eq:eta-fit}. The blue dotted line is the fit for $q = 2$, and the orange dashed line for $q = 1$.}
\label{fig:eta_fit}
\end{figure}

A very interesting result of our simulations is the fact that for models with different orbital separations but the same $v_{\infty}/v_\mathrm{orb}$, we find that a similar specific angular momentum is removed from the system. 
This suggests that the angular momentum loss, as expressed in the parameter $\eta$, may depend primarily on $v_{\infty}/v_\mathrm{orb}$ and
relatively little on other parameters of the system.
The comparison made above with the results of \citet{jahanara2} and \citet{rochester} strengthens this tentative conclusion, and also suggests that $\eta$ may be relatively insensitive to the binary mass ratio. 

We find that the results of all three sets of simulations, after correcting $\eta$ for the spin contribution as described in Section \ref{sec:comparison}, are fairly well described by the following simple function:

\begin{equation} \label{eq:eta-fit}
\eta = \eta_\mathrm{iso} + \frac{1.2 - \eta_\mathrm{iso}}{1 + \left(2.2 \dfrac{v_{\infty}}{v_\mathrm{orb}}\right)^3}.
\end{equation}
This relation is shown in Fig. \ref{fig:eta_fit} for two values of the mass ratio, $q=2$ (corresponding to our simulations and those of \citealp{rochester}) and $q=1$ (corresponding to the results of \citealp{jahanara2}). The function is constructed to converge for very small $v_{\infty}/v_\mathrm{orb}$ at $\eta = 1.2$, which is the maximum value found by \citet{jahanara2} in their low-velocity, `mechanically driven' wind models and appears to be independent of $q$. Most of the points are reasonably well fitted by this relation, the main exception being the \citet{rochester} result for $v_{\infty}/v_\mathrm{orb} \approx 0.8$ which is well above the line. However, the sparse data from hydrodynamical simulations available so far do not warrant a fitting function with a larger number of adjustable parameters. We stress that we consider Equation \ref{eq:eta-fit} to be very preliminary.
A larger grid of simulations with different mass ratios, separations
and mass-loss rates will be needed in order to investigate if in general the specific angular-momentum loss of the material can be written simply in terms of the velocity ratio.

\begin{figure}
\centering
\includegraphics[width=\hsize]{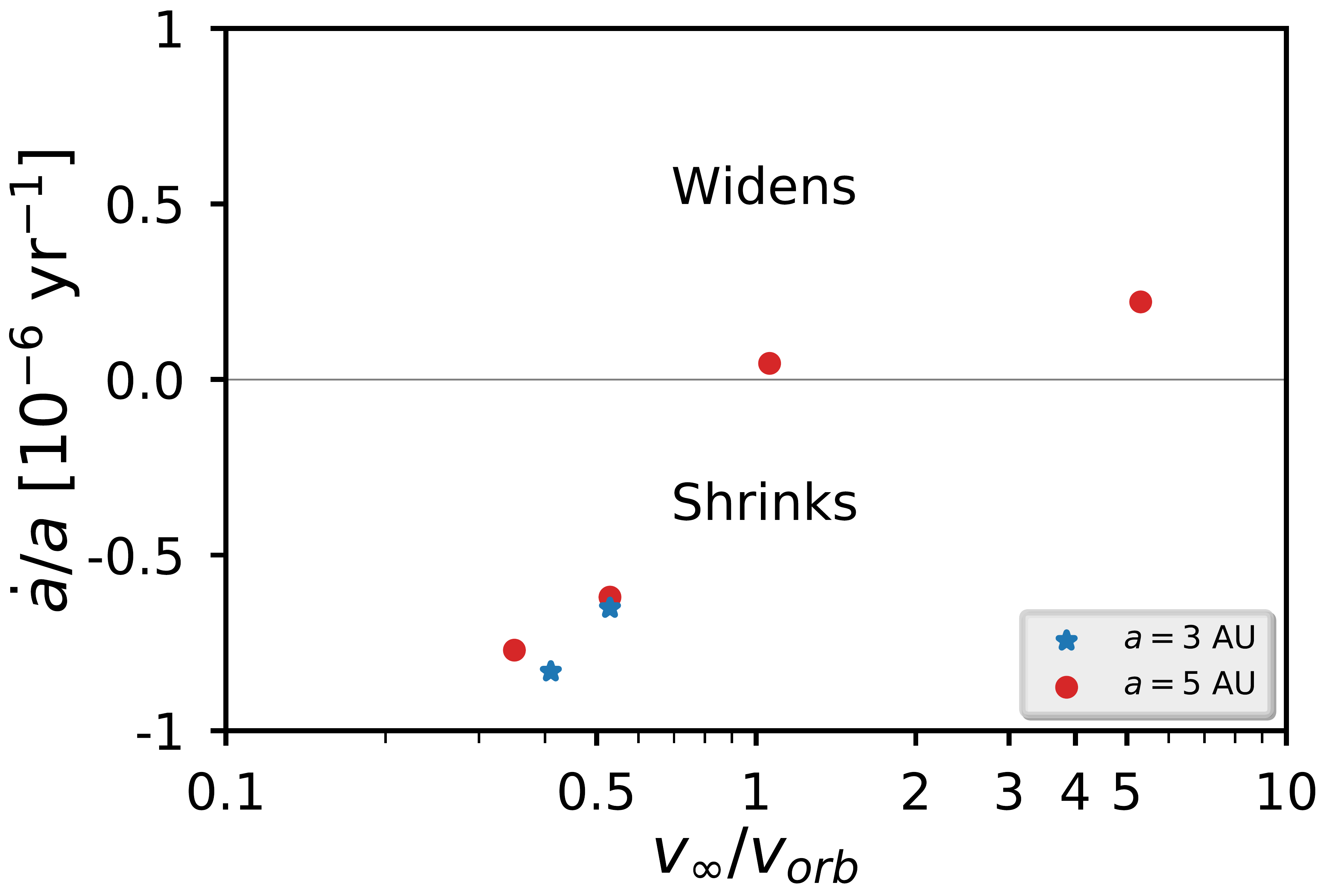}
\caption{The evolution of the orbital separation as a function of $v_{\infty}/v_\mathrm{orb}$ for the systems simulated in this paper. The systems with $\dot{a}/a <0 $ will shrink whereas the ones with $\dot{a}/a>0$ will expand.}
\label{fig:adot}
\end{figure}

Knowing the amount of angular momentum-loss and the fraction of mass accreted by the companion, i.e. the values of $\eta$ and $\beta$, we can predict the evolution of the orbital separation by means of equation
\ref{eq:a}. Figure \ref{fig:adot} shows the theoretical prediction for $\dot{a}/a$ based on the values obtained for $\beta$ and $\eta$ for a mass ratio $q=2$. 
For velocity ratios $v_{\infty}/v_\mathrm{orb} > 1$, we conclude that the orbit should widen, and for high terminal wind velocities relative to the orbital velocity the isotropic-wind regime is approached, with an accretion rate similar to the BHL approximation. 
On the other hand, when $v_{\infty}/v_\mathrm{orb} < 1$ the orbit will shrink, on a timescale similar to the mass-loss timescale of the AGB star. 
As an example, by integrating Eq. 3 over time and keeping constant the value of the mass-loss rate and the values of $\beta$ and $\eta$ obtained for model V15a5, we find that by the time the AGB star reaches a typical WD mass of 0.6 M$_{\odot}$, the orbital separation will shrink by a factor of $\approx 0.6$, from 5~AU to $\approx$ 3~AU. 

This has important consequences for the evolution of AGB binaries and may help to explain the puzzling orbital periods of Ba stars, CEMP-$s$ stars and other post-AGB binary systems as discussed in Sect.~\ref{sec:introduction}. 
\citet{carlo3, carlo2} showed that in order to simultaneously explain the observed abundances and the short orbital periods of individual CEMP-$s$ stars, efficient mass accretion and enhanced angular-momentum loss is needed compared to the predictions of BHL accretion and an isotropic stellar wind. In their models they used a value for the specific angular momentum of the escaping gas equal to two times the average specific angular momentum of the binary, i.e. $\gamma=2$. Using Eq.~\ref{eq:eta-gamma} and a mass ratio $q \approx 2$, this translates into $\eta$ values compatible with our simulations for $v_{\infty}/v_\mathrm{orb} < 1$.
In their population synthesis models of Ba stars \citet{izzard} found that including angular-momentum loss with the same value $\gamma=2$ helps to reproduce the observed period distribution, while the isotropic-wind model produces periods that are too long.
However, the impact of our results, in particular the relation between $\eta$ and the wind velocity ratio, on the period distributions of post-AGB binaries has to be verified by population synthesis modelling. 

Our results also suggest that the number of systems entering a CE phase will increase as a result of the shrinking orbits and high accretion efficiency during low-velocity wind interaction. 
This is important given that many classes of evolved binaries are thought to be the product of a CE phase,
such as cataclysmic binaries consisting of an accreting WD and a Roche-lobe filling low-mass main-sequence star \citep{Knigge+2011}, binary central stars of planetary nebulae \citep{miszalski} and close double white-dwarf binaries \citep{Iben+Tutukov1984}.
All these systems currently have orbital periods ranging from hours to a few days, but must initially have been wide enough to accommodate the red-giant progenitor of a WD, so that drastic orbital shrinkage must have occurred during a CE phase \citep{paczynski}. Furthermore, the double-degenerate formation scenario for type Ia supernovae (SN Ia) invokes the merger of two sufficiently massive carbon-oxygen WDs, brought together into a close orbit during a CE phase and eventually merging due to gravitational-wave emission \citep{iben_tutukov, webbink}. The formation rates of all these systems and events may increase as a result of angular momentum-loss during wind interaction of their wide-orbit progenitors.
In addition, enhanced angular-momentum loss and high accretion efficiencies during wind mass transfer may increase the number of potential SN Ia progenitors via the wide symbiotic channel in the single-degenerate scenario \citep{Hachisu+1999}.

Finally, we note that although our simulations strictly apply to low-mass binaries with AGB donor stars, the same physical processes are likely to occur in other binaries in which a star loses mass via stellar winds. Whenever the wind velocity is similar to or smaller than the orbital velocity, enhanced angular-momentum loss from the orbit may occur. In particular, this may apply to massive binaries under two circumstances: when the mass-losing star is a red supergiant in a very wide binary, or when it is a compact Wolf-Rayet star in a very close binary orbiting a compact object or another Wolf-Rayet star. These binary configurations occur as intermediate stages in several of the progenitor scenarios proposed for the mergers of binary neutron stars and black holes \citep[e.g.][]{Mandel+SdM2016, Tauris+2017, vdHeuvel+2017}. Dedicated simulations of stellar-wind interaction in such massive binaries are needed to quantify the effect on the evolution of their orbits and the possible consequences for the detection rates of gravitational waves by compact binary inspirals \citep{Belczynski+2002, Chruslinska+2018}.

\section{Summary and conclusions}\label{sec:conclude}

We have performed hydrodynamical simulations of low-mass binaries in which one of the components is an AGB star losing mass by a stellar wind at a constant rate and with constant velocity.
The companion star is represented by a sink mass with radius equal to a fraction (0.06-0.1) of its Roche lobe and is located at separations of 3-5 AU from the AGB star. We perform simulations for different ratios of the terminal wind velocity to the relative orbital velocity $v_{\infty}/v_\mathrm{orb}$, in order to study the effect of wind mass loss on the orbits of the binary by determining the specific angular momentum of the mass that is lost and the mass-accretion rate onto the companion. 

We find two regimes of interaction in terms of the $v_{\infty}/v_\mathrm{orb}$ ratio. For cases in which $v_{\infty} < v_\mathrm{orb}$ an accretion disk is formed around the 
companion star as well as two spiral arms around the system that merge at larger distances. The inner spiral arm wraps closely around the mass-losing star and a consists of 
high density gas. In these systems, we also find a large value for the accretion efficiency $\beta$ as well as a high angular-momentum loss per unit mass of material lost from the system. 
The values of both quantities increase with decreasing velocity of the wind relative to the orbital velocity. 
For $v_{\infty} > v_\mathrm{orb}$ the BHL accretion regime is approached and the angular-momentum loss is smaller than for cases with
$v_{\infty} < v_\mathrm{orb}$. In these models, only one or two spiral arms are observed with relatively low gas density. No accretion disk is found in any
of these models; if a disk forms it must be smaller than the assumed sink radius. Our models indicate that the exchange of angular momentum occurs at close distances from the center of mass of the binary and is mainly caused by the torque between the binary
and the gas in the accretion wake that forms behind the accretor. The strength of this torque increases with the gas density in the wake and its misalignment angle, both which are larger for smaller values of $v_{\infty} < v_\mathrm{orb}$.
 
Even though the orbital separations chosen in our simulations are in a regime where wind Roche-lobe overflow is expected, we do not find the characteristic flow through the inner Lagrangian point encountered in WRLOF simulations because we impose a constant-velocity outflow from the AGB star. Nevertheless, we find similarly high accretion efficiencies in our low-wind-velocity models as in WRLOF simulations, in which the gradual acceleration of the wind is modelled explicitly. Our work suggests that gravitational focussing by the companion, in combination with efficient gas cooling, are the main processes that result in a high accretion efficiency during wind mass transfer. We note, however, that accretion rates are difficult to determine accurately from our simulations, in contrast to the angular-momentum loss rates which are computationally very robust.
 
Based on the results we obtain for the mass-accretion efficiency and the specific angular momentum of the material lost, we predict the effect on the orbital separation. We find that for $v_{\infty} < v_\mathrm{orb}$ the orbits will shrink and when $v_{\infty} > v_\mathrm{orb}$ the orbits will widen. 
Furthermore, for the same ratio of wind velocity to orbital velocity we find approximately the same value for the specific angular momentum of the material lost. This suggests that the velocity ratio is the main factor determining the orbital evolution in systems undergoing wind mass transfer.
Our results can help explain the puzzling orbits of post-AGB binaries, such as Ba stars and CEMP-$s$ stars, but this has to be verified by binary population synthesis models. 
Our results also suggest that systems that are initially too wide to undergo Roche-lobe overflow can enter a CE phase, which will have consequences for 
the expected formation rates of systems such as cataclysmic variables, Type Ia supernovae and white-dwarf mergers producing gravitational waves.

\begin{acknowledgements}
We would like to thank the anonymous referee for the very helpful comments in order to improve this paper.
We also would like to thank Carlo Abate, Zhengwei Liu and Richard Stancliffe for the interesting discussions about the method and the results presented in this paper. Additionally we would like to thank Sebastian Ohlmann for discussions on the hydrodynamics method. MIS wants to thank Frank Verbunt, Thomas Wijnen and Andrei Igoshev for the interesting discussions on the topic of this paper. 
\end{acknowledgements}

\bibliographystyle{aa} 
\bibliography{bibliography}

\end{document}